\documentclass[12pt, draftclsnofoot, onecolumn]{IEEEtran}
\usepackage{color}
\usepackage{float}
\usepackage{bm}
\usepackage{amsmath}
\usepackage{amssymb}
\usepackage{graphicx}
\usepackage{cite}
\usepackage{amsfonts}
\usepackage{changepage}
\usepackage{subfigure}
\usepackage{algorithm}
\usepackage{algorithmic}
\usepackage{comment}
\usepackage{multirow}
\usepackage{booktabs}
\def\blue#1{{\color{black}#1}}

\begin{document}
	
\title{Structured Turbo Compressed Sensing for Downlink Massive MIMO-OFDM Channel Estimation}
\author{Xiaoyan~Kuai, Lei~Chen,
	Xiaojun~Yuan,~\IEEEmembership{Senior~Member,~IEEE,}
	and~An~Liu,~\IEEEmembership{Senior~Member,~IEEE}
\thanks{The work of X. Yuan was supported by the China Recruitment Program of Global Young Experts, and the work of A. Liu was supported by the National Science Foundation of China under Project No. 61571383.}
\thanks{X. Kuai and X. Yuan are with the Center for Intelligent Networking and Communications,
the National Laboratory of Science and Technology on Communications, University of Electronic Science and Technology of China, Chengdu 611731, China (e-mail: {xy$\_$kuai, xjyuan}@uestc.edu.cn).}
\thanks{L. Chen is with the Shanghai Institute of Microsystem and Information Technology, Chinese Academy of Sciences, Shanghai 200050, China, University of Chinese Academy of Sciences, Beijing 100049, China, and
  with the School of Information Science and Technology, ShanghaiTech University, Shanghai 201210, China (e-mail: chenlei@shanghaitech.edu.cn).}
\thanks{A. Liu is with the Department of Electronic and Computer Engineering, The Hong Kong University of Science and Technology, Hong Kong (e-mail: eewendaol@ust.hk).}
\thanks{The work has been partially presented in IEEE International Conference on Communications (ICC) 2018 \cite{chen2017massive}.}
}

\markboth{Journal of \LaTeX\ Class Files,~Vol.~XX, No.~X, Month~Year}%
{Shell \MakeLowercase{\textit{et al.}}: Bare Demo of IEEEtran.cls for IEEE Journals}

\maketitle
	
\begin{abstract}
 Compressed sensing has been employed to reduce the pilot overhead for channel estimation in wireless communication systems. Particularly, structured turbo compressed sensing (STCS) provides a generic framework for structured sparse signal recovery with reduced computational complexity and storage requirement. In this paper, we consider the problem of massive multiple-input multiple-output (MIMO) orthogonal frequency division multiplexing (OFDM) channel estimation in a frequency division duplexing (FDD) downlink system. By exploiting the structured sparsity in the \blue{angle-frequency} domain (AFD) and \blue{angle-delay} domain (ADD) of the massive MIMO-OFDM channel, we represent the channel by using AFD and ADD probability models and design message-passing based channel estimators under the STCS framework. Several STCS-based algorithms are proposed for massive MIMO-OFDM channel estimation by exploiting the structured sparsity. We show that, compared with other existing algorithms, the proposed algorithms
 have a much faster convergence speed and achieve  competitive error performance under a wide range of simulation settings.
\end{abstract}


\IEEEpeerreviewmaketitle

\section{Introduction}
Massive multiple-input multiple-output (MIMO) techniques can be combined with orthogonal frequency division multiplexing (OFDM) to achieve huge performance gains in both spectrum and energy efficiency.
 Given that both MIMO and OFDM techniques have been already deployed in the existing commercialized wireless networks, massive MIMO-OFDM has been widely recognized as a high-priority option for future 5G wireless communications \cite{lu2014overview,bolcskei2006mimo,wang2014cellular}.

In massive MIMO-OFDM, the acquisition of accurate channel state information (CSI) is essential for harvesting the capacity and reliability enhancement promised by the system. However, conventional channel estimation approaches require that the pilot length should be at least the same as the number of transmit antennas\blue{\cite{yuan2015fundamental,hassibi2003much}}. This may cause a significant pilot overhead in the downlink of a massive MIMO-OFDM system, where a large number of antennas are deployed at base station (BS). A possible solution to this problem is to assume time division duplexing (TDD), where the CSI only needs to be acquired in the uplink and then the downlink CSI is automatically obtained thanks to the channel reciprocity. However, on one hand, due to limited coherence time and the mismatch of uplink and downlink transmit-receive filters, the CSI acquired in the uplink might be inaccurate for the downlink transmission. On the other hand, it is economically disadvantageous to deploy TDD systems since frequency division duplexing (FDD) dominates the current cellular networks \cite{Gao_TCOM2016_csMIMO}. Therefore, it is of critical importance to reduce the pilot overhead for the downlink of FDD massive MIMO-OFDM systems.

Due to limited local \blue{scatterers} in physical environments, a massive MIMO-OFDM channel usually exhibits abundant sparsity in certain transformed domains \cite{bajwa2010compressed,berger2010application}. Compressed sensing (CS) algorithms, such as orthogonal matching pursuit (OMP) \cite{tropp2007signal} and least absolute shrinkage and selection operator (LASSO) algorithm \cite{berger2010application}, have recently been used to exploit the sparsity in the channel estimation of massive MIMO-OFDM, so as to reduce the pilot overhead. In particular, a burst LASSO algorithm is developed in \cite{liu2016exploiting} for clustering the non-zero channel coefficients in the virtual angle domain of a massive MIMO channel. \blue{By exploiting temporal correlation of a massive MIMO channel, the CS algorithms in \cite{XRVL15} and \cite{HLDavid17} further reduce the pilot overhead.}
In \cite{gao2015spatially}, an algorithm named distributed sparsity adaptive matching pursuit (DSAMP) was proposed to jointly estimate the channel coefficients of multiple subcarriers based on the common sparsity in the frequency domain. 
 In \cite{Gao_TCOM2016_csMIMO}, channel estimation is designed on account of the temporal correlation of the sparsity in the delay domain.

However, the above CS-based channel estimation algorithms, when applied to FDD downlink massive MIMO-OFDM, have their respective drawbacks. For example, OMP and LASSO do not take into account the structured sparsity in the algorithm design. The burst LASSO algorithm, which considers group sparsity in the virtual angle domain, only works well when non-zero clusters of the channel coefficients have a similar size. The DSAMP algorithm considers the common sparsity between different subcarriers, but unfortunately does not exploit the sparse structure in the angle domain.

Recently, message-passing algorithms for compressed sensing \blue{\cite{donoho2009message,vila2013expectation,Ma2014Turbo,Ma2015On,xue2017denoising,chen2017,MSRJr18,ziniel2013dynamic,LWJKYH18}}
have attracted much research interest due to their fast convergence and low computational complexity.
Among them, the turbo compressed sensing (Turbo-CS) algorithm \cite{Ma2014Turbo} and its variants \cite{xue2017denoising,chen2017} have the state-of-the-art performance in both complexity and convergence rate, especially when partial orthogonal sensing matrices are involved. In particular, the authors in \cite{chen2017} proposed a modified Turbo-CS algorithm, termed structured turbo compressed sensing (STCS), in which Turbo-CS is combined with a Markov model to efficiently exploit the clustered sparsity of the massive MIMO channel in the angle domain.
\blue{In this paper, the main contributions of our work are summarized as follows.
\begin{itemize}
  \item We extend STCS for the channel estimation of massive MIMO-OFDM by exploiting
  structured channel sparsity not only in the angle-frequency domain, but also in the angle-delay domain\footnote{\blue{The structured compressed sensing algorithm exploits the sparsity structure in
the angle-delay domain of the massive MIMO-OFDM system, while the structured compressed sensing algorithm in \cite{Gao_TCOM2016_csMIMO}
exploits the spatial temporal common sparsity of the MIMO-OFDM system.}}.
  \item We develop a Markov chain as the probability model to characterize the structured sparsity  of massive MIMO-OFDM channels
  in the angle-frequency domain. The resulting algorithm is referred to as STCS with frequency support (STCS-FS).
  We further develop a Markov model with two types of hidden state variables to describe the structured sparsity of the massive MIMO-OFDM channel in the angle-delay domain.
  The resulting algorithm is referred to as STCS with delay support (STCS-DS).
 We develop state evolution (SE) to accurately predict the performance of the STCS-FS and STCS-DS algorithms.
  \item Extensive simulation results are presented to show the advantages of the STCS-FS and STCS-DS algorithms.
In particular, we show that STCS-DS exhibits the fastest convergence rate among all the existing CS-based
iterative algorithms, and achieves competitive mean-square error (MSE) performance.
Both algorithms have been validated under realistic channel models.
\end{itemize}
}
\blue{\subsection{Related Work}}

\blue{In the recent work \cite{ziniel2013dynamic}, the dynamic compressive sensing (CS)
  problem of recovering sparse, correlated, time-varying signals
  from sub-Nyquist, non-adaptive, linear measurements was explored
  from a Bayesian perspective. Compared to \cite{ziniel2013dynamic}, the novelty of our work is as follows.
  First, we aim to extend STCS for the channel estimation of massive MIMO-OFDM by exploiting structured channel sparsity in the
  angle-delay domain, while \cite{ziniel2013dynamic} aims to solve the dynamic CS problem of recovering sparse, correlated, time-varying signals.
   Second, the probability process in this paper is used to model the clustering property of the channel coefficient support
   in the angle-frequency domain and angle-delay domain, while the Markov process is used to model the time-varying coefficient
   support and the time-varying coefficient amplitudes in \cite{ziniel2013dynamic}.
   Third, the Turbo-CS algorithm is used in our paper, while approximate message passing (AMP) was used to recover the sparse signal in \cite{ziniel2013dynamic}.
    The turbo-CS algorithm has lower complexity and exhibits faster convergence speed than the AMP algorithm \cite{xue2017denoising}. }

\blue{Compared with the recent work \cite{LWJKYH18}, the novelty of our work consists of the following aspects.
First, we employ a Markov model to efficiently exploit the clustered sparsity of the
massive MIMO channel in the angle-frequency domain and angle-delay domain, while \cite{LWJKYH18} uses
the nearest neighbor sparsity pattern learning (NNSPL) algorithm first proposed in \cite{MWKHL16} to exploit the sparsity structure.\footnote{\blue{Note that
\cite{LWKNMJ17} uses the NNSPL algorithm to exploit the angle domain sparsity of the channel, while \cite{WNMK16} uses the NNSPL
algorithm to exploit the delay domain sparsity. \cite{LWJKYH18} presents a comprehensive version of the NNSPL algorithm to jointly handle
the angle-delay domain sparsity.}}
Second, STCS-FS achieves a considerably lower mean square error (MSE) performance than the NNSPL algorithm with frequency support,
  while STCS-DS performs slightly better than the NNSPL algorithm with delay support.
  Third, the computational complexity of STCS is much lower than that of NNSPL.
  In this regard, we show that the per-iteration complexity of STCS is lower than that of
  NNSPL; we further show that STCS converges much faster than NNSPL.}
\subsection{Organization}
The rest of the paper is organized as follows. In Section \ref{section:sysmod}, we present the massive MIMO-OFDM channel model and the joint sparsity in the \blue{angle-frequency} and \blue{angle-delay} domain. In Section \ref{section:PreSTCS}, we review the STCS framework. In Sections \ref{section:comsup} and \ref{section:bitsup}, the details of the angle-frequency domain and angle-delay domain channel support models are elaborated. Based on these models, we design the STCS-based algorithms. Simulation results and conclusions are presented in Sections \ref{section:simcomp} and \ref{section:conclu}, respectively.

\section{System Model}
\label{section:sysmod}
\subsection{Massive MIMO-OFDM}
Consider a typical massive MIMO-OFDM system with one BS serving multiple single-antenna users, where the BS comprises $N$ antennas and the system employs $P$ pilot subcarriers. Without loss of generality, we focus on the downlink channel estimation problem at a reference user. To estimate the downlink channel $\bm{\tilde{h}}^{(p)}_{f} \in \mathbb{C}^{N \times 1}$ of the reference user at the $p$-th pilot subcarrier in the frequency domain, the BS sends $M$ training symbols $\bm{x}^{(p)}_{m} \in \mathbb{C}^{N \times 1}$, $1 \leq m \leq M$ over successive time slots. Then the received signal at the reference user $\bm{y}^{(p)}_{f} \in \mathbb{C}^{M \times 1}$ can be written as
\begin{equation}
	\bm{y}^{(p)}_{f} = \bm{X}^{(p)}\bm{\tilde{h}}^{(p)}_{f} + \bm{w}^{(p)}_{f}, 1 \leq p \leq P,
	\label{equ:spatfre}
\end{equation}
where $\bm{X}^{(p)} = [\bm{x}^{(p)}_{1},\cdots,\bm{x}^{(p)}_{M}]^{\text{T}}$ is an $M \times N$ pilot matrix, and $\bm{w}^{(p)}_{f} \sim \mathcal{CN}(\bm{0},\sigma^{2}\bm{I})$ is an additive white Gaussian noise (AWGN).
From \cite{Tse:2005:FWC:1111206,Gao_TCOM2016_csMIMO,liu2016exploiting}, $\bm{\tilde{h}}^{(p)}_{f}$ is sparse in the angle domain, i.e., $\bm{\tilde{h}}^{(p)}_{f}$ can be expressed as
\begin{equation}
	\bm{\tilde{h}}^{(p)}_{f} = \bm{B}\bm{h}^{(p)}_{f}, 1 \leq p \leq P,
	\label{equ:chantrans}
\end{equation}
where $\bm{B}$ is the transform matrix determined by the geometrical structure of the antenna array, and $\bm{h}^{(p)}_{f}$ is a sparse representation of the channel in the transform domain\footnote{The sparsity of $\bm{h}_{f}^{(p)}$ is due to a limited number of scatterers at the BS in a typical wireless environment.}. In this paper, we focus on the half-wavelength uniform linear array (ULA) at BS, where $\bm{B}$ is the inverse discrete Fourier transform (DFT) matrix \cite{Tse:2005:FWC:1111206}. Our discussion can be readily extended to a uniform planar array (UPA) or higher-dimensional antenna array. Substituting (\ref{equ:chantrans}) and letting $\bm{A}^{(p)} = \bm{X}^{(p)}\bm{B}$, we can rewrite (\ref{equ:spatfre}) as
\begin{equation}
	\bm{y}^{(p)}_{f} = \bm{A}^{(p)}\bm{h}^{(p)}_{f} + \bm{w}^{(p)}_{f}, 1 \leq p \leq P.
	\label{equ:angfre2}
\end{equation}
Our goal is to estimate the sparse vectors $\bm{h}^{(p)}_{f}$ from the low-dimensional observed signal $\bm{y}^{(p)}_{f}$, $1 \leq p \leq P$. This problem can be solved by the existing compressed sensing algorithms \cite{berger2010application,liu2016exploiting,gao2015spatially,Gao_TCOM2016_csMIMO,Ma2014Turbo,donoho2009message,vila2013expectation,chen2017}. However, these existing algorithms, if directly applied, can not efficiently exploit the unique sparsity structure of the massive MIMO-OFDM channel, as detailed below.

\subsection{Channel Sparsity}
\label{section:chanspar}
\begin{figure}[htbp]
	\centering{}
	\includegraphics[width=3.2in]{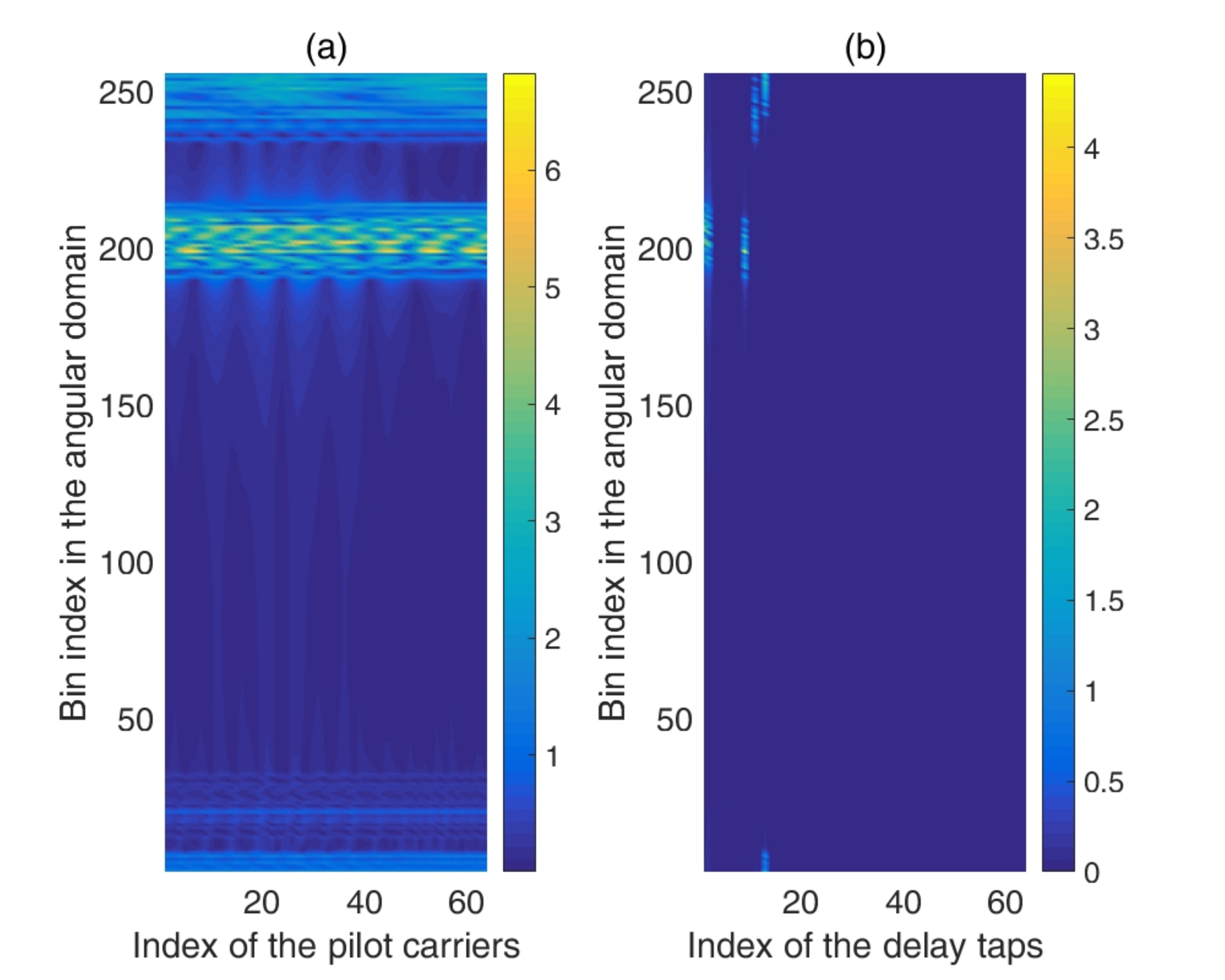}
	\caption{An example of the spatial channel model in \cite{WinnerScmImplementationIEEETranbst}. Subfigure (a) plots the channel gains in the angle-frequency domain, and subfigure (b) plots the channel gains in the angle-delay domain. All parameters in the model are set by default, except for \text{'NumBsElements'} = 256, \text{'NumMsElements'} = 1, and \text{'Scenario'} = Urban-macro.}
	\label{fig:SCMChan}
\end{figure}

Due to the scattering effect, a massive MIMO-OFDM channel exhibits clustered sparsity. Besides, the \blue{scatterers} for different subchannels are very similar \cite{Tse:2005:FWC:1111206}. Consequently, for a communication system with the bandwidth much smaller than the carrier frequency (e.g., \blue{10~MHz} in LTE-A systems with a carrier frequency of \blue{2~GHz}), the subchannels $\{ \bm{h}^{(p)}_{f} \}_{p=1}^{P}$ have a common support for sparsity \cite{gao2015spatially}, i.e.
\begin{equation}
	\text{supp}\{\bm{h}^{(1)}_{f}\} = \text{supp}\{\bm{h}^{(2)}_{f}\} = \cdots = \text{supp}\{\bm{h}^{(P)}_{f}\},
\end{equation}
where \text{supp}$\{\bm{h}_{f}^{(p)}\}$ returns the positions of the non-zero entries of $\bm{h}_{f}^{(p)}$.
As an example, in Fig. \ref{fig:SCMChan} (a), we generate a massive MIMO-OFDM channel using the spatial channel model (SCM) \cite{WinnerScmImplementationIEEETranbst} with carrier frequency at \blue{2~GHz}, bandwidth \blue{7.5~MHz} and frequency interval \blue{15~kHz}. There are 512 subcarriers in total, and 64 of them are chosen as pilot subcarriers. It is clear that massive MIMO-OFDM subchannels have a common support in the frequency domain, and the non-zero elements appear in a clustered manner in the angle domain. 

A limited number of \blue{scatterers} also cause sparsity in the delay domain \cite{barhumi2003optimal,Gao_TCOM2016_csMIMO}. We can transform the channel response matrix from the \blue{angle-frequency} domain to the \blue{angle-delay} domain with an inverse Fourier transform \cite{dai2013spectrally,barhumi2003optimal}, i.e.,
\begin{equation}
\bm{H}_{f}\bm{F}^{\ast} = \bm{H}_{d},
\label{equ:dftrans}
\end{equation}
where $\bm{H}_{f} = [\bm{h}_{f}^{(1)},\cdots,\bm{h}_{f}^{(P)}]$, $\bm{H}_{d} = [\bm{h}_{d}^{(1)},\cdots,\bm{h}_{d}^{(P)}]$, $\bm{F}$ denotes the $P \times P$ unitary DFT matrix, and $(\cdot)^{\ast}$ denotes the conjugate operation. Without loss of generality, let $L$ be the maximum delay spread, implying $\bm{h}_{d}^{(p)} = \bm{0}$ for $p = L+1, \cdots, P$. Fig. \ref{fig:SCMChan} (b) shows the channel matrix $\bm{H}_{d}$ in the angle-delay domain. We see that many columns of $\bm{H}_{d}$ approach zero, and the non-zero elements are grouped into a small number of clusters.

\section{Structured Turbo Compressed Sensing}
\begin{figure}[htbp]
	\centering{}
	\includegraphics[width=3.2in]{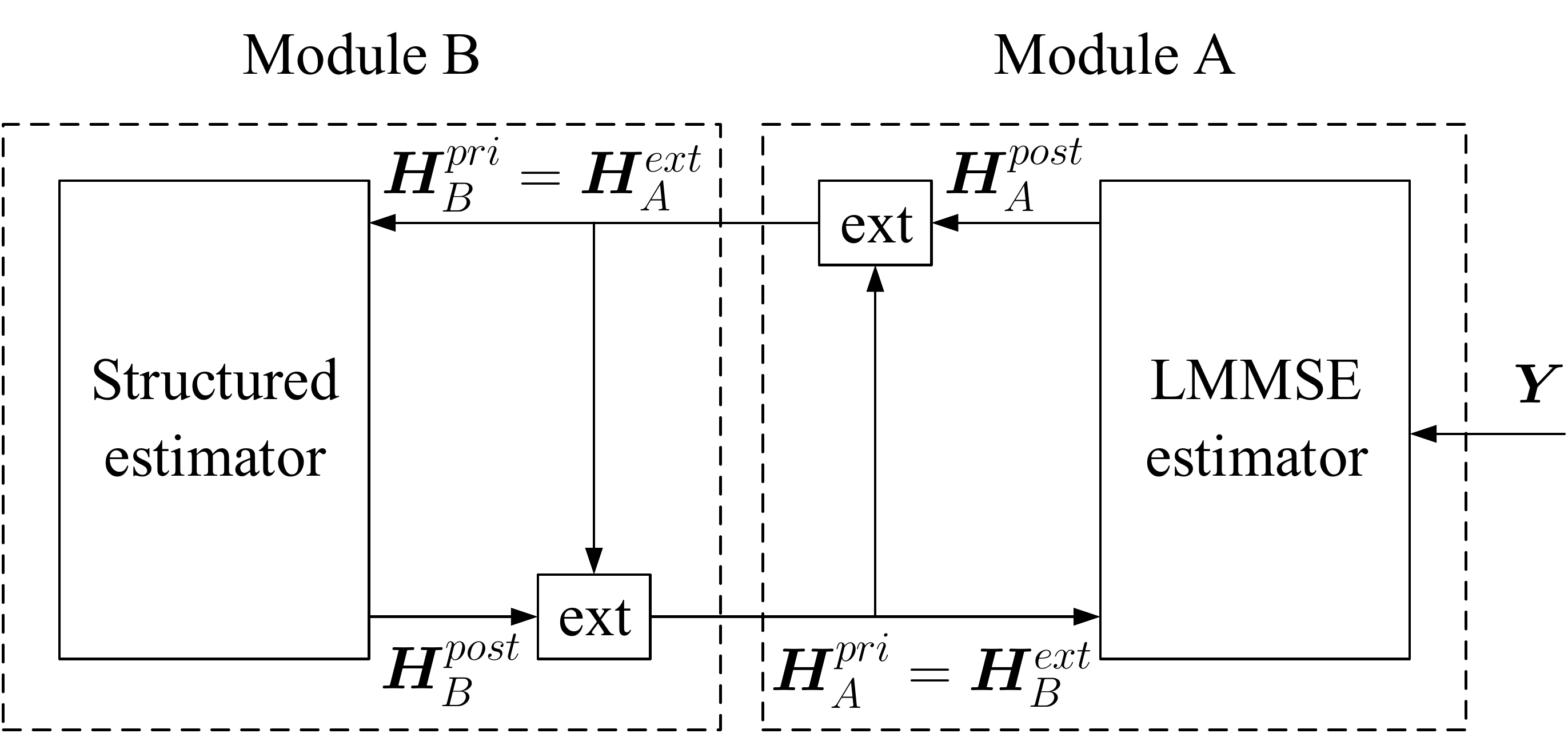}
	\caption{\label{fig:StrTurboCSdiag} An illustration of the structured Turbo-CS framework.}
\end{figure}
\label{section:PreSTCS}

The goal of this paper is to estimate $\bm{H}_{f}$ based on the observed signal $\bm{Y}_{f} = [\bm{y}_{f}^{(1)}, \bm{y}_{f}^{(2)}, \cdots, \bm{y}_{f}^{(P)}]$ together with the sparsity of $\bm{H}_{f}$ described in Section $\ref{section:chanspar}$. We will mainly follow the STCS approach in \cite{Ma2014Turbo,xue2017denoising,chen2017} to solve the above problem. For self-containedness, we present the STCS algorithm in the following. For notational convenience, we drop the subscripts of $\bm{H}_{f}$ and $\bm{Y}_{f}$, since later we will apply the STCS algorithm to the delay domain representations of $\bm{H}_{f}$ and $\bm{Y}_{f}$.

The STCS algorithm contains two modules, namely, Module A and Module B. Module A is basically a linear minimum mean square error (LMMSE) estimator based on the observation $\bm{Y}$ and the messages from Module B. Module B refines the estimate of the channel by combining the messages from Module A and the prior distribution of $\bm{H}$. The two modules are executed iteratively until convergence.

In Module A, the channel vector $\bm{h}^{(p)}$ (the $p$-th column of matrix $\bm{H}$) is estimated based on the observation $\bm{y}^{(p)}$ (the $p$-th column of matrix $\bm{Y}$) with a prior distribution $\mathcal{CN}(\bm{h}^{(p)};\bm{h}_{A}^{pri(p)},v_{A}^{pri(p)}\bm{I})$\footnote{ $\mathcal{CN}(x;\bar{x},\sigma_{x}^{2})$ is a complex Gaussian distribution of $x$ with mean $\bar{x}$ and variance $\sigma^{2}_{x}$. Similar notations are used for $\bm{H}_{A}^{post}$, $\bm{H}_{A}^{ext}$, $\bm{H}_{B}^{pri}$, $\bm{H}_{B}^{post}$ and $\bm{H}_{B}^{ext}$.}, where $\bm{h}_{A}^{pri(p)}$ is the $p$-th column of $\bm{H}_{A}^{pri}$ and $v_{A}^{pri(p)}$ is the corresponding variance. Note that $\bm{h}_{A}^{pri(p)}$ and $v_{A}^{pri(p)}$ are the extrinsic mean and variance from Module B. Then the posterior distribution of $\bm{h}^{(p)}$ is still complex Gaussian with mean and variance given by
\begin{equation}
	\bm{h}_{A}^{post(p)}\!=\!\bm{h}_{A}^{pri(p)} + \frac{v_{A}^{pri(p)}}{v_{A}^{pri(p)}+\sigma^{2}}\bm{A}^{(p)H}\left(\bm{y}^{(p)}\!\!-\!\!\bm{A}^{(p)}\bm{h}_{A}^{pri(p)}\right),
\end{equation}
and
\begin{equation}
	v_{A}^{post(p)} = v_{A}^{pri(p)} - \frac{M}{N}\frac{(v_{A}^{pri(p)})^{2}}{v_{A}^{pri(p)}+\sigma^{2}},
\end{equation}
for all $p$. Then we need to calculate the extrinsic messages following the message passing principle \cite{berrou1996near}. The extrinsic distribution of $\bm{h}^{(p)}$ satisfies
\begin{equation}
\begin{aligned}
	&\mathcal{CN}(\bm{h}^{(p)};\bm{h}_{A}^{post(p)},v_{A}^{post(p)}\bm{I})\\
	&\propto \mathcal{CN}(\bm{h}^{(p)};\bm{h}_{A}^{pri(p)},v_{A}^{pri(p)}\bm{I})\mathcal{CN}(\bm{h}^{(p)};\bm{h}_{A}^{ext(p)},v_{A}^{ext(p)}\bm{I}).\\
\end{aligned}
\label{equ:extexplain}
\end{equation}
Then the extrinsic mean and variance are given by
\begin{equation}
	\bm{h}_{B}^{pri(p)} = \bm{h}_{A}^{ext(p)} = v_{A}^{ext(p)}\left(\frac{\bm{h}_{A}^{post(p)}}{v_{A}^{post(p)}} - \frac{\bm{h}_{A}^{pri(p)}}{v_{A}^{pri(p)}}\right),
	\label{equ:exthb}
\end{equation}
and
\begin{equation}
	v_{B}^{pri(p)} = v_{A}^{ext(p)} = \left(\frac{1}{v_{A}^{post(p)}}-\frac{1}{v_{A}^{pri(p)}}\right)^{-1}.
	\label{equ:extvb}
\end{equation}

The key challenge resides in the design of Module B. Specifically, we need to design a structured estimator that can efficiently exploit both the angle domain sparsity and the delay-domain sparsity of the massive MIMO-OFDM channel. We will develop probability models to describe the sparsity structure of the MIMO-OFDM channel. Based on that, we construct factor graphs and design message passing algorithms for the realization of Module B. The details are presented in Sections \ref{section:comsup} and \ref{section:bitsup}.

\section{Frequency Domain Channel Support Model}
\label{section:comsup}
\subsection{Probability Model}
From Section \ref{section:sysmod}, the channel coefficients for the massive MIMO-OFDM system cluster in the angle domain and share a common support in the frequency domain. Motivated by this observation, we use a Markov chain with a common hidden variable to describe such a channel structure. The probability model can be written as 
\begin{equation}
p(h_{f,n}^{(p)}|s_{f,n}) = (1-s_{f,n})\delta(h_{f,n}^{(p)}) + s_{f,n}G(h_{f,n}^{(p)}),
\label{equ:comsupmodel}
\end{equation}
where $s_{f,n} \in \{0, 1\}$ is a hidden binary state indicating whether the channel coefficients $h_{f,n}^{(p)}$ for all $p$ are all zero $(s_{f,n} = 0)$ or not $(s_{f,n} = 1)$, $\delta(\cdot)$ is the Dirac function, and $G(\cdot)$ denotes the probability distribution for non-zero coefficients. In this paper, function $G$ is chosen as $\mathcal{CN}(h_{f,n}^{(p)};0,(\sigma_{f}^{(p)})^{2})$. Define a vector $\bm{s}_{f} = [s_{f,1},\cdots,s_{f,N}]^{T}$. Then the clustering effect of non-zeros can be modeled using a Markov chain as
\begin{equation}
p(\bm{s}_{f}) = p(s_{f,1})\prod_{n=2}^{N}p(s_{f,n}|s_{f,n-1}),
\end{equation}
with the transition and initial probabilities given by
\begin{equation}
p(s_{f,n}|s_{f,n-1})=
\left\{
\begin{aligned}
(1-p_{10})^{1-s_{f,n}} (p_{10})^{s_{f,n}},\quad & s_{f,n-1}=0;\\
(p_{01})^{1-s_{f,n}} (1-p_{01})^{s_{f,n}},\quad & s_{f,n-1}=1\\
\end{aligned}
\right.
\end{equation}
and
\begin{equation}
p(s_{f,1}) = (1-\lambda_{f})^{1-s_{f,1}}(\lambda_{f})^{s_{f,1}},
\end{equation}
where $\lambda_{f} \triangleq \text{Pr}\{s_{f,n} = 1\} = (1+p_{01}/p_{10})^{-1}$, the average ratio of the non-zero elements in $\bm{s}_{f}$, describes the sparsity of $\bm{h}_{f}^{(p)}$ for all $p$. Such a Markov chain is fully described by parameters $p_{10} = \text{Pr}(s_{f,n}=1|s_{f,n-1}=0)$ and $p_{01} = \text{Pr}(s_{f,n}=0|s_{f,n-1}=1)$. 
\blue{Since $p_{00}=1-p_{10}$, a smaller $p_{10}$ implies a larger gap between two clusters. Similarly, with $p_{11}=1-p_{01}$, a smaller $p_{01}$ implies a larger average cluster size.}

\subsection{Message Passing for Module B}
\begin{figure}[htbp]
	\centering{}
	\includegraphics[width=3in]{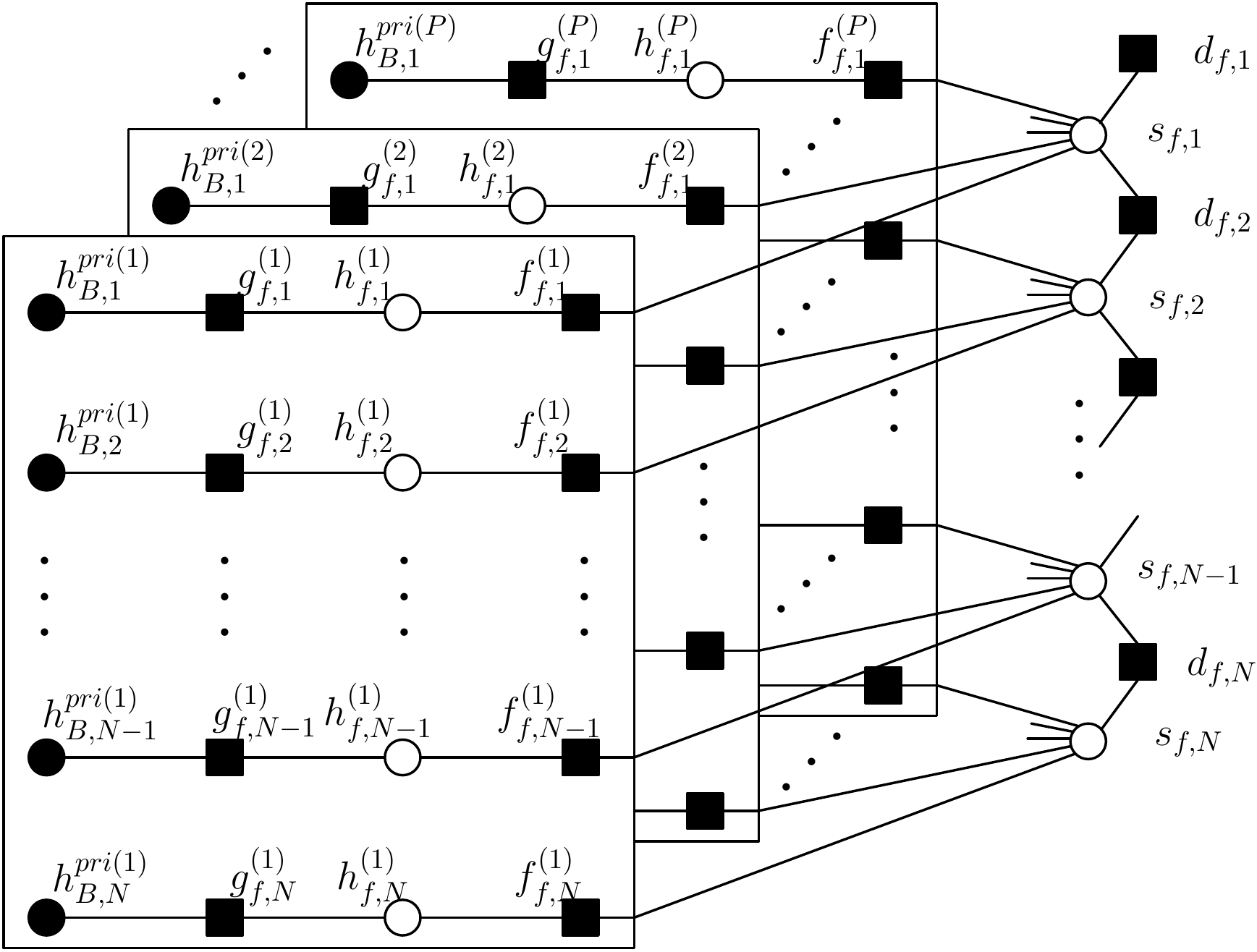}
	\caption{\label{fig:ComSup} Factor graph $\mathcal{G}_{f}$ for the frequency support channel model.}
\end{figure}
\begin{table*}[htbp]
	\centering
	\caption{Functions of the factor nodes in Fig. \ref{fig:ComSup}}
	\label{tab1}
	\begin{tabular}{cc}
		\hline
		Factor Node& Factor Function \\
		\hline
		$g_{f,n}^{(p)}$ & $g_{f,n}^{(p)}(h_{B,n}^{pri(p)},h_{f,n}^{(p)}) = p(h_{B,n}^{pri(p)}|h_{f,n}^{(p)}) = \mathcal{CN}(h_{f,n}^{(p)};h_{B,n}^{pri(p)},v_{B}^{pri(p)})$\\
		$f_{f,n}^{(p)}$ & $f_{f,n}^{(p)}(h_{f,n}^{(p)},s_{f,n}) = p(h_{f,n}^{(p)}|s_{f,n}) = (1-s_{f,n})\delta(h_{f,n}^{(p)}) + s_{f,n}\mathcal{CN}( h_{f,n}^{(p)};0,(\sigma_{f}^{(p)})^{2})$\\
		$d_{f,1}$ & $d_{f,1}(s_{f,1}) = p(s_{f,1}) = (1-\lambda_{f})^{1-s_{f,1}}(\lambda_{f})^{s_{f,1}}$\\
		$d_{f,n}$ & $d_{f,n}(s_{f,n},s_{f,n-1}) = p(s_{f,n}|s_{f,n-1}) =  \left\{
		\begin{aligned}
		(1-p_{10})^{1-s_{f,n}} (p_{10})^{s_{f,n}},\quad & s_{f,n-1}=0\\
		(p_{01})^{1-s_{f,n}} (1-p_{01})^{s_{f,n}},\quad & s_{f,n-1}=1\\
		\end{aligned}
		\right.$\\
 		\hline
	\end{tabular}
\end{table*}

In this subsection, we explain the details of Module B for the frequency support model in the angle-frequency domain. First of all, a basic assumption is used to model $\bm{h}_{B}^{pri(p)}$, the input mean of Module B in (\ref{equ:exthb}), as
\begin{equation}
	\bm{h}_{B}^{pri(p)} = \bm{h}_{f}^{(p)} + \bm{n}_{f}^{(p)}, \forall p,
	\label{equ:assump1}
\end{equation}
where $\bm{n}_{f}^{(p)} \sim \mathcal{CN}(\bm{0},v_{B}^{pri(p)}\bm{I})$ is independent of $\bm{h}_{f}^{(p)}$, and $v_{B}^{pri(p)}$ is the input variance of Module B in (\ref{equ:extvb}).
  Similar assumptions have been used in message-passing-based iterative signal recovery algorithm \cite{donoho2009message, vila2013expectation, Ma2014Turbo, ma2017orthogonal, xue2017denoising, chen2017}. Under this assumption, the factor graph of the joint probability distribution
\begin{equation}
\begin{aligned}
	&p(\bm{H}_{B}^{pri},\bm{H}_{f},\bm{s}_{f})\\
	&= p(\bm{H}_{B}^{pri}|\bm{H}_{f})p(\bm{H}_{f}|\bm{s}_{f})p(\bm{s}_{f})\\
	&= \prod_{p=1}^{P}\prod_{n=1}^{N}p(h_{B,n}^{pri(p)}|h_{f,n}^{(p)})\prod_{p=1}^{P}\prod_{n=1}^{N}p(h_{f,n}^{(p)}|s_{f,n})\\
	&\qquad\qquad\cdot p(s_{f,1})\prod_{n=2}^{N}p(s_{f,n}|s_{f,n-1}),\\
\end{aligned}
\end{equation}
denoted by $\mathcal{G}_{f}$, is shown in Fig. \ref{fig:ComSup}, where the factor function of each factor node is listed in Table \ref{tab1}.

We now give a message passing algorithm based on graph $\mathcal{G}_{f}$. According to the sum-product rule, the message from variable node $h_{f,n}^{(p)}$ to factor node $f_{f,n}^{(p)}$ is
\begin{equation}
	\nu_{h_{f,n}^{(p)}\rightarrow f_{f,n}^{(p)}}(h_{f,n}^{(p)})=\mathcal{CN}(h_{f,n}^{(p)};h_{B,n}^{pri(p)},v_{B}^{pri(p)}),
	\label{equ:hf2ff}
\end{equation}
and the message from factor node $f_{f,n}^{(p)}$ to variable node $s_{f,n}$ is
\begin{equation}
\begin{aligned}
&\nu_{f_{f,n}^{(p)}\rightarrow s_{f,n}}(s_{f,n})\\
&\propto \int_{h_{f,n}^{(p)}} f_{f,n}^{(p)}(h_{f,n}^{(p)},s_{f,n}) \cdot \nu_{h_{f,n}^{(p)}\rightarrow f_{f,n}^{(p)}}\\
& = \stackrel{\rightharpoonup}{\pi}_{n}^{(p)}s_{f,n}+(1-\stackrel{\rightharpoonup}{\pi}_{n}^{(p)})(1-s_{f,n}),
\end{aligned}
\end{equation}
where $f_{f,n}^{(p)}$ is given by Table \ref{tab1}, and
\begin{equation}
	\stackrel{\rightharpoonup}{\pi}_{n}^{(p)} = \left(1+\frac{\mathcal{CN}(0;h_{B,n}^{pri(p)},v_{B}^{pri(p)})}{\mathcal{CN}(0;h_{B,n}^{pri(p)},v_{B}^{pri(p)} + (\sigma_{f}^{(p)})^{2})}\right)^{-1}.
\end{equation}

Then forward-backward message passing is performed over the binary Markov chain $\bm{s}_{f}$, with the forward and backward messages respectively given by
\begin{equation}
\begin{aligned}
&\nu_{d_{f,n} \rightarrow s_{f,n}}\\
&\propto\!\!\sum_{s_{f,n-1}} \!\!\!d_{f,n}(s_{f,n},s_{f,n-1})\nu_{d_{f,n-1} \rightarrow s_{f,n-1}}\prod\limits_{p=1}^{P}\nu_{f_{f,n-1}^{(p)}\rightarrow s_{f,n-1}}\\
&= \lambda_{n}^{f}s_{f,n} + (1-\lambda_{n}^{f})(1-s_{f,n}),
\end{aligned}
\end{equation}
and
\begin{equation}
\begin{aligned}
&\nu_{d_{f,n+1} \rightarrow s_{f,n}} \\
&\propto\!\!\sum_{s_{f,n+1}} \!\!\!d_{f,n}(s_{f,n+1},s_{f,n})\nu_{d_{f,n+2} \rightarrow s_{f,n+1}}\prod\limits_{p=1}^{P}\nu_{f_{f,n+1}^{(p)}\rightarrow s_{f,n+1}}\\
&= \lambda_{n}^{b}s_{f,n} + (1-\lambda_{n}^{b})(1-s_{f,n}),
\end{aligned}
\end{equation}
where
\begin{equation}
	\lambda_{n}^{f} =
	\frac{(1-p_{01})\frac{\lambda_{n-1}^{f}}{1-\lambda_{n-1}^{f}}\prod\limits_{p=1}^{P}\frac{\stackrel{\rightharpoonup}{\pi}_{n-1}^{(p)}}{1-\stackrel{\rightharpoonup}{\pi}_{n-1}^{(p)}} + p_{10} }{\frac{\lambda_{n-1}^{f}}{1-\lambda_{n-1}^{f}}\prod\limits_{p=1}^{P}\frac{\stackrel{\rightharpoonup}{\pi}_{n-1}^{(p)}}{1-\stackrel{\rightharpoonup}{\pi}_{n-1}^{(p)}} + 1},
\end{equation}
and
\begin{equation}
	\lambda_{n}^{b} = \frac{ (1-p_{01})\frac{\lambda_{n+1}^{b}}{1-\lambda_{n+1}^{b}}\prod\limits_{p=1}^{P}\frac{\stackrel{\rightharpoonup}{\pi}_{n+1}^{(p)}}{1-\stackrel{\rightharpoonup}{\pi}_{n+1}^{(p)}} + p_{01}}{ (1-p_{01}+p_{10})\frac{\lambda_{n+1}^{b}}{1-\lambda_{n+1}^{b}}\prod\limits_{p=1}^{P}\frac{\stackrel{\rightharpoonup}{\pi}_{n+1}^{(p)}}{1-\stackrel{\rightharpoonup}{\pi}_{n+1}^{(p)}} + (1-p_{10}+p_{01})},
\end{equation}
with $\lambda_{1}^{f} = \lambda_{f}$ and $\lambda_{N}^{b} = 1/2$.

After that, according to the sum-product rule, the message from variable node $s_{f,n}$ to factor node $f_{f,n}^{(p)}$ is
\begin{equation}
\begin{aligned}
&\nu_{s_{f,n} \rightarrow f_{f,n}^{(p)}}(s_{f,n})\\
&\propto \nu_{d_{f,n} \rightarrow s_{f,n}}\nu_{d_{f,n+1} \rightarrow s_{f,n}}\prod\limits_{p'\neq p}^{}\nu_{f_{f,n}^{(p')}\rightarrow s_{f,n}}\\
&=\stackrel{\leftharpoonup}{\pi}_{n}^{(p)}s_{f,n}+(1-\stackrel{\leftharpoonup}{\pi}_{n}^{(p)})(1-s_{f,n}),
\end{aligned}
\end{equation}
where
\begin{equation}
	\stackrel{\leftharpoonup}{\pi}_{n}^{(p)}=\frac{ \lambda_{n}^{f}\lambda_{n}^{b}\prod\limits_{p' \neq p}\stackrel{\rightharpoonup}{\pi}_{n}^{(p')} }{ (1-\lambda_{n}^{f})(1-\lambda_{n}^{b})\prod\limits_{p' \neq p}(1-\stackrel{\rightharpoonup}{\pi}_{n}^{(p')} )+\lambda_{n}^{f}\lambda_{n}^{b}\prod\limits_{p' \neq p}\stackrel{\rightharpoonup}{\pi}_{n}^{(p')} }.
\end{equation}
The message from the factor node $f_{f,n}^{(p)}$ back to variable node $h_{f,n}^{(p)}$ is
\begin{equation}
\begin{aligned}
	&\nu_{f_{f,n}^{(p)}\rightarrow h_{f,n}^{(p)}}(h_{f,n}^{(p)}) \\
	&\propto\sum_{s_{f,n}} f_{f,n}^{(p)}(h_{f,n}^{(p)},s_{f,n}) \cdot \nu_{s_{f,n}\rightarrow f_{f,n}^{(p)}}\\
	&=\stackrel{\leftharpoonup}{\pi}_{n}^{(p)}\mathcal{CN}(h_{f,n}^{(p)};0,(\sigma_{f}^{(p)})^{2})+(1-\stackrel{\leftharpoonup}{\pi}_{n}^{(p)})\delta(h_{f,n}^{(p)}).
\end{aligned}
\end{equation}
\begin{algorithm}
	\caption{\label{alg1}Structured Turbo-CS Algorithm with Frequency Support (STCS-FS)}
	\begin{algorithmic}
		
		\REQUIRE received signal $\bm{Y}_{f} = [\bm{y}^{(1)}_{f}, \cdots, \bm{y}^{(P)}_{f}]$, pilot matrix $\bm{X}^{(p)}$ $\forall p$, and additive noise variance $\sigma^{2}$.
		
		\ENSURE channel state information $\bm{\hat{H}}$.
		
		\textbf{Initialize:} $\bm{A}^{(p)}=\bm{X}^{(p)}\bm{F}^{H}$, $\bm{h}_{A}^{pri(p)}$, $v_{A}^{pri(p)}$, $\forall p$.
		
		\textbf{Module A:}
		
		\textbf{\% LMMSE estimator}
		
		1: $\bm{h}_{A}^{post(p)}\!=\!\bm{h}_{A}^{pri(p)}+\frac{v_{A}^{pri(p)}}{v_{A}^{pri(p)}+\sigma^{2}}\bm{A}^{(p)H}(\bm{y}_{f}^{(p)}-\bm{A}^{(p)}\bm{h}_{A}^{pri(p)})$, $\forall p$.
		
		2: $v_{A}^{post(p)}=v_{A}^{pri(p)}-\frac{M}{N}\cdot\frac{(v_{A}^{pri(p)})^{2}}{v_{A}^{pri(p)}+\sigma^{2}}$, $\forall p$.
		
		\textbf{\% Update extrinsic messages}
		
		3: $v_{B}^{pri(p)}=v_{A}^{ext(p)}=\left(\frac{1}{v_{A}^{post(p)}}-\frac{1}{v_{A}^{pri(p)}}\right)^{-1}$, $\forall p$.
		
		4: $\bm{h}_{B}^{pri(p)}=\bm{h}_{A}^{ext(p)}=v_{B}^{pri(p)}\left(\frac{\bm{h}_{A}^{post(p)}}{v_{A}^{post(p)}}-\frac{\bm{h}_{A}^{pri(p)}}{v_{A}^{pri(p)}}\right)$, $\forall p$.
		
		\textbf{Module B:}
		
		\textbf{\% Structured estimator}
		
		5: $\bm{h}_{B}^{post(p)}$ and $v_{B}^{post(p)}$, $\forall p$, are given by (\ref{equ:postmea}) and (\ref{equ:postvar}).
		
		\textbf{\% Update extrinsic messages}
		
		6: $v_{A}^{pri(p)}=v_{B}^{ext(p)}=\left(\frac{1}{v_{B}^{post(p)}}-\frac{1}{v_{B}^{pri(p)}}\right)^{-1}$, $\forall p$.
		
		7: $\bm{h}_{A}^{pri(p)}=\bm{h}_{B}^{ext(p)}=v_{A}^{pri(p)}\left(\frac{\bm{h}_{B}^{post(p)}}{v_{B}^{post(p)}}-\frac{\bm{h}_{B}^{pri(p)}}{v_{B}^{pri(p)}}\right)$, $\forall p$.
		
		Repeat Module A and Module B until convergence or the maximum iteration number is exceeded.
	\end{algorithmic}
\end{algorithm}

The posterior mean and variance can be calculated as
\begin{equation}
	h_{B,n}^{post(p)} = \text{E}(h_{f,n}^{(p)}|\bm{H}_{B}^{pri}) = \int_{h_{f,n}^{(p)}} h_{f,n}^{(p)}p(h_{f,n}^{(p)}|\bm{H}_{B}^{pri}),
	\label{equ:postmea}
\end{equation}
and
\begin{equation}
\begin{aligned}
	v_{B}^{post(p)} &= \frac{1}{N}\sum_{n=1}^{N}\text{Var}(h_{f,n}^{(p)}|\bm{H}_{B}^{pri})\\
	&= \frac{1}{N}\sum_{n=1}^{N}\int_{h_{f,n}^{(p)}}|h_{f,n}^{(p)} - h_{B,n}^{post(p)}|^{2}p(h_{f,n}^{(p)}|\bm{H}_{B}^{pri}),\\
	\label{equ:postvar}
\end{aligned}
\end{equation}
where the conditional distribution is
\begin{equation}
	p(h_{f,n}^{(p)}|\bm{H}_{B}^{pri}) \propto \nu_{g_{f,n}^{(p)} \rightarrow h_{f,n}^{(p)}}\nu_{f_{f,n}^{(p)} \rightarrow h_{f,n}^{(p)}},
\end{equation}
with $\nu_{g_{f,n}^{(p)} \rightarrow h_{f,n}^{(p)}} = \mathcal{CN}(h_{f,n}^{(p)};h_{B,n}^{pri(p)},v_{B}^{pri(p)})$.

Based on the derivation in \cite{Ma2014Turbo,xue2017denoising}, the corresponding extrinsic update can be calculated as
\begin{equation}
	v_{A}^{pri(p)} = v_{B}^{ext(p)} = \left( \frac{1}{v_{B}^{post(p)}} - \frac{1}{v_{B}^{pri(p)}} \right)^{-1},
	\label{equ:extvar}
\end{equation}
and
\begin{equation}
	\bm{h}_{A}^{pri(p)} = \bm{h}_{B}^{ext(p)} = v_{A}^{pri(p)}\left( \frac{\bm{h}_{B}^{post(p)}}{v_{B}^{post(p)}} - \frac{\bm{h}_{B}^{pri(p)}}{v_{B}^{pri(p)}}\right).
	\label{equ:extmea}
\end{equation}
The structured Turbo-CS algorithm with Module B realized by Eqs. (\ref{equ:hf2ff}) to (\ref{equ:extmea}) is referred to as structured Turbo-CS with frequency support (STCS-FS). The STCS-FS algorithm is summarized in Algorithm \ref{alg1}.

\section{Delay Domain Channel Support Model}
\label{section:bitsup}
\subsection{Probability Model}
In this section, we establish a probability model to directly characterize the angle-delay domain sparsity.
To start with, we transform the system model in (\ref{equ:angfre2}) into the angle-delay domain.
\blue{Define
\begin{align}
\bm{Y}_{f}&=[\bm{y}^{(1)}_{f},\cdots,\bm{y}^{(P)}_{f}] \notag \\
\bm{H}_{f}&=[\bm{h}^{(1)}_{f},\cdots,\bm{h}^{(P)}_{f}] \notag \\
\bm{W}_{f}&=[\bm{w}^{(1)}_{f},\cdots,\bm{w}^{(P)}_{f}]. \notag
\end{align}
As the pilot matrix can be designed in advance, for simplification,
we assume that $\mathbf{X}^{(1)}=\cdots=\mathbf{X}^{(P)}$, i.e., the matrix $\bm{X}$ is the same for
different pilot subcarrier. $\mathbf{X}^{(p)}$ is hence replaced by $\mathbf{X}$ for $ 1 \leq p \leq P$.
The received signal in the angle-frequency domain is represented as
\begin{equation}
\bm{Y}_{f}=\bm{A}\bm{H}_{f}+\bm{W}_{f}.
\end{equation}
We next transform the channel response matrix from the angle-frequency domain to the angle-delay domain with an inverse
Fourier transform $\bm{F}^{*}$, i.e., $\bm{H}_{f}\bm{F}^{*}=\bm{H}_d$ in \eqref{equ:dftrans}.
Then the received signal in the delay domain can be represented as
\begin{align}\label{eq.delay.matrix1}
\bm{Y}_{f}\bm{F}^{*}&=\bm{A}\bm{H}_{f}\bm{F}^{*}+\bm{W}_{f}\bm{F}^{*},
\end{align}
or equivalently,
\begin{align}\label{system.DS}
\bm{Y}_{d}&=\bm{A}\bm{H}_d+\bm{W}_d.
\end{align}
From \eqref{system.DS}, the $p$-th column of $\bm{Y}_{d}$ is given by}
\begin{equation}
\bm{y}_{d}^{(p)} = \bm{A}\bm{h}_{d}^{(p)} + \bm{w}_{d}^{(p)}, 1 \leq p \leq P,
\label{equ:angdel}
\end{equation}
where $\bm{y}_{d}^{(p)}$ is the received signal in the delay domain, $\bm{w}_{d}^{(p)} \sim \mathcal{CN}(\bm{0},\sigma^{2}\bm{I})$ is an AWGN with the same variance as $\bm{w}_{f}^{(p)}$, and $\bm{h}_{d}^{(p)}$ is the channel coefficient vector in the delay domain\footnote{\blue{The STCS algorithms in this paper can be extended
to the system model such as (13) in \cite{MSRJr18}. However, this involves more complicated signal processing since then the
path delay taps are mixed in the channel output.}}.

In the angle-delay domain, two hidden binary states are introduced to model the non-zero columns and cluster structure of the delay domain channel matrix $\bm{H}_{d}$. Each channel coefficient $h_{d,n}^{(p)}$ has a conditionally independent distribution expressed as
\begin{equation}
	\begin{aligned}
	p&(h_{d,n}^{(p)}|s_{d,n}^{(p)})\\
	&=(1-s_{d,n}^{(p)})\delta(h_{d,n}^{(p)}) + s_{d,n}^{(p)}\mathcal{CN}( h_{d,n}^{(p)};0,(\sigma_{d}^{(p)})^{2} ),\\
	\end{aligned}
	\label{equ:bitsupmodel}
\end{equation}
where $s_{d,n}^{(p)} \in \{0,1\}$ is a hidden binary state. What is different from the angle-frequency model is that the binary state $s_{d,n}^{(p)}$ is also conditioned on another binary state $t_{d}^{(p)} \in \{0,1\}$. Specifically, $t_{d}^{(p)}$ indicates whether the $p$-th column of $\bm{H}_{d}$ is zero $(t_{d}^{(p)} = 0)$ or not $(t_{d}^{(p)} = 1)$. Hence, the vector $\bm{t}_{d} = [t_{d}^{(1)},\cdots,t_{d}^{(P)}]$ can be used to capture the channel sparsity in the delay domain and each entry $t_{d}^{(p)}$ complies with a Bernoulli distribution
\begin{equation}
p(t_{d}^{(p)}) = (1-\gamma_{d}^{(p)})^{1-t_{d}^{(p)}}(\gamma_{d}^{(p)})^{t_{d}^{(p)}},
\end{equation}
where $\gamma_{d}^{(p)}$ denotes the probability that the $p$-th column in $\bm{H}_{d}$ is non-zero. On the other hand, $s_{d,n}^{(p)}$ indicates whether the $(n,p)$-th element of $\bm{H}_{d}$ is zero $(s_{d,n}^{(p)} = 0)$ or not $(s_{d,n}^{(p)} = 1)$. Hence, the vector $\bm{s}_{d}^{(p)} = [s_{d,1},\cdots,s_{d,N}^{(p)}]^{T}$ can be used to capture the clustered sparsity in the angle domain for the channel vector $\bm{h}^{(p)}_{d}$. Specifically, conditioned on $t^{(p)}_{d}$, the cluster structure of $\bm{h}^{(p)}_{d}$ can be modeled using a Markov chain as
\begin{equation}
p(\bm{s}_{d}^{(p)}|t_{d}^{(p)}) = p(s_{d,1}^{(p)}|t_{d}^{(p)})\prod_{n=2}^{N}p(s_{d,n}^{(p)}|s_{d,n-1}^{(p)},t_{d}^{(p)}),
\end{equation}
with the transition and initial probabilities given by
\begin{equation}
\begin{aligned}
p(&s_{d,n}^{(p)}|s_{d,n-1}^{(p)},t_{d}^{(p)})\\
&=
\left\{
\begin{aligned}
(1-p_{10}^{(p)})^{1-s_{d,n}^{(p)}} (p_{10}^{(p)})^{s_{d,n}^{(p)}},\quad & s_{d,n-1}^{(p)}=0, t_{d}^{(p)}=1;\\
(p_{01}^{(p)})^{1-s_{d,n}^{(p)}} (1-p_{01}^{(p)})^{s_{d,n}^{(p)}},\quad & s_{d,n-1}^{(p)}=1, t_{d}^{(p)}=1;\\
1^{1-s_{d,n}^{(p)}}0^{s_{d,n}^{(p)}},\quad & s_{d,n-1}^{(p)}=0, t_{d}^{(p)}=0;\\
1^{1-s_{d,n}^{(p)}}0^{s_{d,n}^{(p)}},\quad & s_{d,n-1}^{(p)}=1, t_{d}^{(p)}=0\\
\end{aligned}
\right.\\
\end{aligned}
\end{equation}
and
\begin{equation}
\begin{aligned}
&p(s_{d,1}^{(p)}|t^{(p)}_{d})=\\
(1&-\lambda_{d}^{(p)})^{(1-s_{d,1}^{(p)})t_{d}^{(p)}}(\lambda_{d}^{(p)})^{s_{d,1}^{(p)}t_{d}^{(p)}}1^{(1-s_{d,1}^{(p)})(1-t_{d}^{(p)})}0^{s_{d,1}^{(p)}(1-t_{d}^{(p)})}\!\!.\\
\end{aligned}
\end{equation}
In other words, when $t^{(p)}_{d} = 0$, we must have $\bm{s}_{d}^{(p)} = \bm{0}$. When $t^{(p)}_{d} = 1$, $\bm{s}_{d}^{(p)}$ is a binary Markov chain similar to $\bm{s}_{f}$ but with different transition probabilities. The probability model is illustrated as a factor graph in Fig. \ref{fig:BitSup}.

\subsection{Message Passing for Module B}

\begin{figure}[htbp]
	\centering{}
	\includegraphics[width=3in]{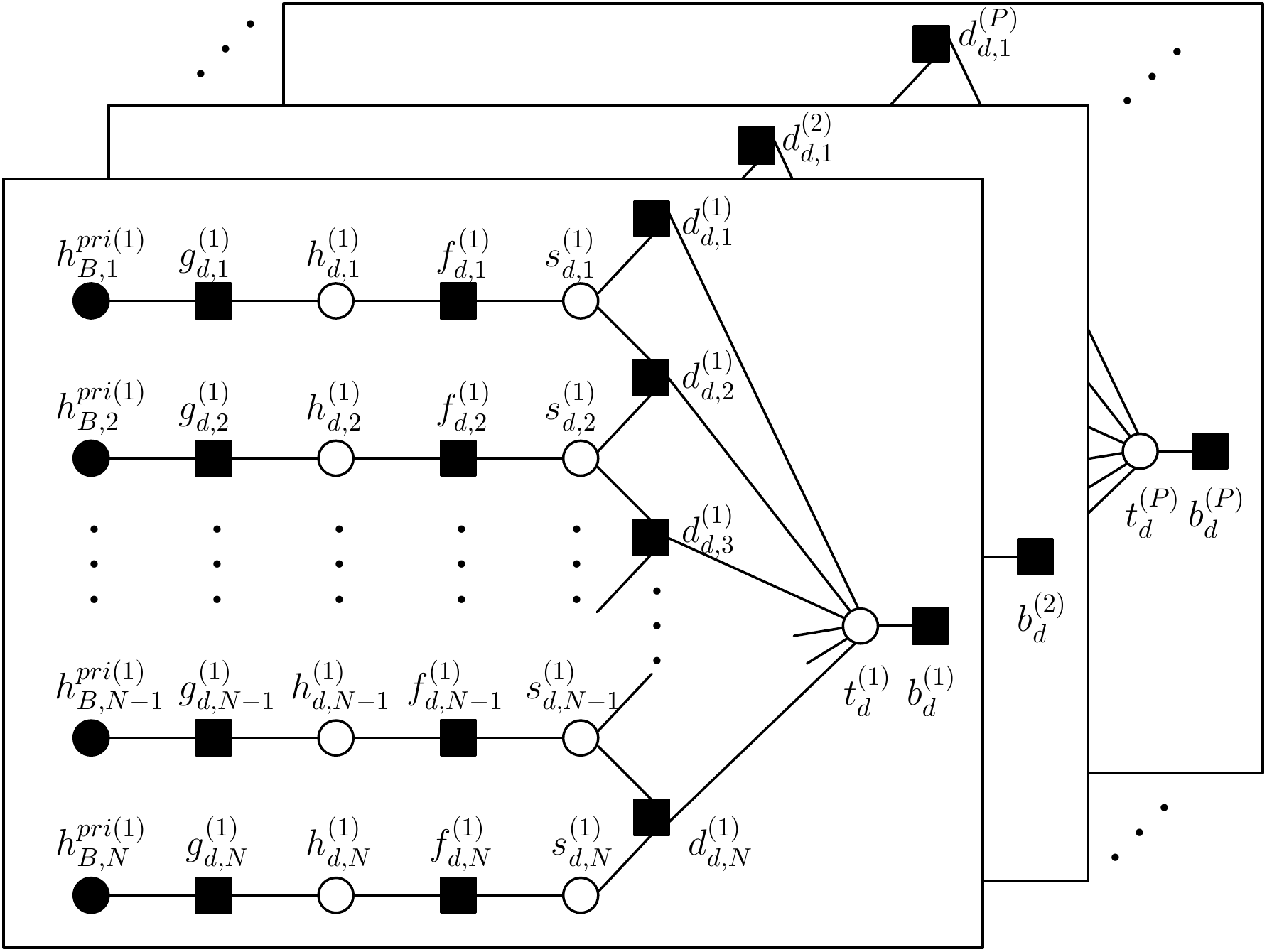}
	\caption{\label{fig:BitSup} Factor graph $\mathcal{G}_{d}$ for the delay support model.}
\end{figure}
\begin{table*}[htbp]
	\centering
	\caption{Functions of the factor nodes in Fig. \ref{fig:BitSup}}
	\label{tab2}
	\begin{tabular}{cc}
		\hline
		Factor Node & Factor Function\\
		\hline
		$g_{d,n}^{(p)}$ & $g_{d,n}^{(p)}(h_{B,n}^{pri(p)},h_{d,n}^{(p)}) = p(h_{B,n}^{pri(p)}|h_{d,n}^{(p)}) = \mathcal{CN}(h_{d,n}^{(p)};h_{B,n}^{pri(p)},v_{B}^{pri(p)})$\\
		$f_{d,n}^{(p)}$ & $f_{d,n}^{(p)}(h_{d,n}^{(p)},s_{d,n}^{(p)}) = p(h_{d,n}^{(p)}|s_{d,n}^{(p)}) = (1-s_{d,n}^{(p)})\delta(h_{d,n}^{(p)}) + s_{d,n}^{(p)}\mathcal{CN}( h_{d,n}^{(p)};0,(\sigma_{d}^{(p)})^{2})$\\
		$d_{d,1}^{(p)}$ & $d_{d,1}^{(p)}(s_{d,1}^{(p)}, t_{d}^{(p)}) = p(s_{d,1}^{(p)}|t_{d}^{(p)}) = (1-\lambda_{d}^{(p)})^{(1-s_{d,1}^{(p)})t_{d}^{(p)}}(\lambda_{d}^{(p)})^{s_{d,1}^{(p)}t_{d}^{(p)}}1^{(1-s_{d,1}^{(p)})(1-t_{d}^{(p)})}0^{s_{d,1}^{(p)}(1-t_{d}^{(p)})}$\\
		$d_{d,n}^{(p)}$ & $d_{d,n}^{(p)}(s_{d,n}^{(p)}, s_{d,n-1}^{(p)}, t_{d}^{(p)}) = p(s_{d,n}^{(p)}|s_{d,n-1}^{(p)},t_{d}^{(p)}) = \left\{
		\begin{aligned}
		(1-p_{10}^{(p)})^{1-s_{d,n}^{(p)}} (p_{10}^{(p)})^{s_{d,n}^{(p)}},\quad & s_{d,n-1}^{(p)}=0, t_{d}^{(p)} = 1\\
		(p_{01}^{(p)})^{1-s_{d,n}^{(p)}} (1-p_{01})^{s_{d,n}},\quad & s_{d,n-1}^{(p)}=1, t_{d}^{(p)} = 1\\
		1^{1-s_{d,n}^{(p)}}0^{s_{d,n}^{(p)}},\quad & s_{d,n-1}^{(p)}=0, t_{d}^{(p)} = 0\\
		1^{1-s_{d,n}^{(p)}}0^{s_{d,n}^{(p)}},\quad & s_{d,n-1}^{(p)}=1, t_{d}^{(p)} = 0\\
		\end{aligned}
		\right.$\\
		$b_{d}^{(p)}$ & $b_{d}^{(p)}(t_{d}^{(p)}) = p(t_{d}^{(p)}) = (1-\gamma_{d}^{(p)})^{1-t_{d}^{(p)}}(\gamma_{d}^{(p)})^{t_{d}^{(p)}}$\\
		\hline
	\end{tabular}
\end{table*}

Similarly to (\ref{equ:assump1}), we assume
\begin{equation}
\bm{h}_{B}^{pri(p)} = \bm{h}_{d}^{(p)} + \bm{n}_{d}^{(p)}, \forall p,
\label{equ:assump2}
\end{equation}
where $\bm{n}_{d}^{(p)} \sim \mathcal{CN}(\bm{0},v_{B}^{pri(p)}\bm{I})$ is independent of $\bm{h}_{d}^{(p)}$. The factor graph of the joint distribution
\begin{equation}
\begin{aligned}
&p(\bm{H}_{B}^{pri},\bm{H}_{d},\bm{S}_{d}, \bm{t}_{d})\\
&= \prod_{p=1}^{P}p(\bm{h}_{B}^{pri(p)},\bm{h}_{d}^{(p)},\bm{s}_{d}^{(p)}, t_{d}^{(p)})\\
&= \prod_{p=1}^{P}\Big[\prod_{n=1}^{N}p(h_{B,n}^{pri(p)}|h_{d,n}^{(p)})\prod_{n=1}^{N}p(h_{d,n}^{(p)}|s_{d,n}^{(p)})\\
&\qquad\cdot p(s_{d,1}^{(p)}|t_{d}^{(p)})\prod_{n=2}^{N}p(s_{d,n}^{(p)}|s_{d,n-1}^{(p)},t_{d}^{(p)})p(t_{d}^{(p)})\Big],\\
\label{equ:promoddel}
\end{aligned}
\end{equation}
denoted by $\mathcal{G}_{d}$, is shown in Fig. \ref{fig:BitSup}, where the function of each factor node is listed in Table \ref{tab2}.
The probability model in (\ref{equ:promoddel}) assumes that the columns of $\bm{H}_{d}$ are independent of each other. This is justified by the fact that in practical scenarios, the channel coefficients for different delay taps usually experience significantly different channel fading.

We now derive the message passing algorithm on graph $\mathcal{G}_{d}$. Note that the functions of $d_{d,1}^{(p)}(s_{d,1}^{(p)}, t_{d}^{(p)})$ and $d_{d,n}^{(p)}(s_{d,n}^{(p)}, s_{d,n-1}^{(p)}, t_{d}^{(p)})$ are modified by replacing $1$ and $0$ with $1-\varepsilon$ and $\varepsilon$, where $\varepsilon$ is a small constant. Such a modification is used in \cite{ziniel2013dynamic} to avoid improper probability distribution functions and make the algorithm more robust.

We start with message passing from variable node $t_{d}^{(p)}$ to factor node $d_{d,1}^{(p)}(s_{d,1}^{(p)}, t_{d}^{(p)})$ or $d_{d,n}^{(p)}(s_{d,n}^{(p)}, s_{d,n-1}^{(p)}, t_{d}^{(p)})$:
\begin{equation}
	\begin{aligned}
	\nu_{t_{d}^{(p)} \rightarrow d_{d,n}^{(p)}} &\propto \nu_{b_{d}^{(p)} \rightarrow t_{d}^{(p)}} \prod_{n' \neq n}\nu_{d_{d,n'}^{(p)} \rightarrow t_{d}^{(p)}}\\
	&\propto \left[ \gamma_{d}^{(p)}t_{d}^{(p)} + (1 - \gamma_{d}^{(p)})(1-t_{d}^{(p)}) \right] \\
	& \qquad \cdot \prod_{n' \neq n} \left[ \stackrel{\rightharpoonup}{\theta}_{n'}^{(p)}t_{d}^{(p)} + (1-\stackrel{\rightharpoonup}{\theta}_{n'}^{(p)})(1-t_{d}^{(p)}) \right] \\
	&= \stackrel{\leftharpoonup}{\theta}_{n}^{(p)}t_{d}^{(p)} + (1-\stackrel{\leftharpoonup}{\theta}_{n}^{(p)})(1-t_{d}^{(p)}),\\
	\end{aligned}
	\label{equ:td2dd}
\end{equation}
where
\begin{equation}
	\stackrel{\leftharpoonup}{\theta}_{n}^{(p)} = \frac{\gamma_{d}^{(p)}\prod\limits_{n' \neq n}\stackrel{\rightharpoonup}{\theta}_{n'}^{(p)}}{\gamma_{d}^{(p)}\prod\limits_{n' \neq n}\stackrel{\rightharpoonup}{\theta}_{n'}^{(p)} + (1-\gamma_{d}^{(p)})\prod\limits_{n' \neq n}(1-\stackrel{\rightharpoonup}{\theta}_{n'}^{(p)}) }.
\end{equation}

The message from factor node $f_{d,n}^{(p)}$ to variable node $s_{d,n}^{(p)}$ is given by
\begin{equation}
\begin{aligned}
\nu_{f_{d,n}^{(p)}\rightarrow s_{d,n}^{(p)}}&(s_{d,n}^{(p)}) =\\
&\propto\int_{h_{d,n}^{(p)}}f_{d,n}^{(p)}(h_{d,n}^{(p)},s_{d,n}^{(p)})\cdot\nu_{h_{d,n}^{(p)}\rightarrow f_{d,n}^{(p)}}\\
&=\stackrel{\rightharpoonup}{\pi}_{n}^{(p)}s_{d,n}^{(p)}+(1-\stackrel{\rightharpoonup}{\pi}_{n}^{(p)})(1-s_{d,n}^{(p)}).\\
\end{aligned}
\end{equation}

Then the forward-backward message passing can be applied in the Markov chains according to the sum-product rule. Note that $s_{d,n}^{(p)}$ and $t_{d}^{(p)}$ are binary variables. The messages passed between $\{s_{d,n}^{(p)}\}$ and $\{t_{d}^{(p)}\}$ are given by
\begin{equation}
	\nu_{d_{d,n}^{(p)} \rightarrow s_{d,n}^{(p)}} = \lambda_{n}^{f(p)}s_{d,n}^{(p)} + (1-\lambda_{n}^{f(p)})(1-s_{d,n}^{(p)}),
\end{equation}
\begin{equation}
	\nu_{s_{d,n}^{(p)} \rightarrow d_{d,n+1}^{(p)}} = \lambda_{n}^{f'(p)}s_{d,n}^{(p)} + (1-\lambda_{n}^{f'(p)})(1-s_{d,n}^{(p)}),
\end{equation}
\begin{equation}
	\nu_{d_{d,n+1}^{(p)} \rightarrow s_{d,n}^{(p)}} = \lambda_{n}^{b(p)}s_{d,n}^{(p)} + (1-\lambda_{n}^{b(p)})(1-s_{d,n}^{(p)}),
\end{equation}
and
\begin{equation}
\nu_{s_{d,n}^{(p)} \rightarrow d_{d,n}^{(p)}} = \lambda_{n}^{b'(p)}s_{d,n}^{(p)} + (1-\lambda_{n}^{b'(p)})(1-s_{d,n}^{(p)}).
\end{equation}
\begin{algorithm}
	\caption{\label{alg3}Structured Turbo-CS Algorithm with Delay Support (STCS-DS)}
	\begin{algorithmic}
		
		\REQUIRE received signal $\bm{Y}_{f} = [\bm{y}^{(1)}_{f}, \cdots, \bm{y}^{(P)}_{f}]$, pilot matrix $\bm{X}^{(p)}$ $\forall p$, and additive noise variance $\sigma^{2}$.
		
		\ENSURE channel state information $\bm{\hat{H}}$.
		
		\textbf{Initialize:} $\bm{Y}_{d}=\bm{Y}_{f}\bm{F}^{\ast}$, $\bm{A}=\bm{X}\bm{F}^{H}$, $\bm{h}_{A}^{pri(p)}$, $v_{A}^{pri(p)}$, $\forall p$.
		
		\textbf{Module A:}
		
		\textbf{\% LMMSE estimator}
		
		1: $\bm{h}_{A}^{post(p)}\!=\!\bm{h}_{A}^{pri(p)}+\frac{v_{A}^{pri(p)}}{v_{A}^{pri(p)}+\sigma^{2}}\bm{A}^{H}(\bm{y}_{d}^{(p)}-\bm{A}\bm{h}_{A}^{pri(p)})$, $\forall p$
		
		2: $v_{A}^{post(p)}=v_{A}^{pri(p)}-\frac{M}{N}\cdot\frac{(v_{A}^{pri(p)})^{2}}{v_{A}^{pri(p)}+\sigma^{2}}$, $\forall p$
		
		\textbf{\% Update extrinsic messages}
		
		3: $v_{B}^{pri(p)}=v_{A}^{ext(p)}=\left(\frac{1}{v_{A}^{post(p)}}-\frac{1}{v_{A}^{pri(p)}}\right)^{-1}$, $\forall p$
		
		4: $\bm{h}_{B}^{pri(p)}=\bm{h}_{A}^{ext(p)}=v_{B}^{pri(p)}\left(\frac{\bm{h}_{A}^{post(p)}}{v_{A}^{post(p)}}-\frac{\bm{h}_{A}^{pri(p)}}{v_{A}^{pri(p)}}\right)$, $\forall p$
		
		\textbf{Module B:}
		
		\textbf{\% Structured estimator}
		
		5: $\bm{h}_{B}^{post(p)}$ and $v_{B}^{post(p)}$, $\forall p$, are given by (\ref{equ:postmea2}) and (\ref{equ:postvar2}).
		
		\textbf{\% Update extrinsic messages}
		
		6: $v_{A}^{pri(p)}=v_{B}^{ext(p)}=\left(\frac{1}{v_{B}^{post(p)}}-\frac{1}{v_{B}^{pri(p)}}\right)^{-1}$, $\forall p$
		
		7: $\bm{h}_{A}^{pri(p)}=\bm{h}_{B}^{ext(p)}=v_{A}^{pri(p)}\left(\frac{\bm{h}_{B}^{post(p)}}{v_{B}^{post(p)}}-\frac{\bm{h}_{B}^{pri(p)}}{v_{B}^{pri(p)}}\right)$, $\forall p$
		
		Repeat Module A and Module B until convergence or the maximum iteration number is exceeded.
	\end{algorithmic}
\end{algorithm}

After that, we calculate the messages going out of the Markov chains and the message back to factor node $f_{d,n}^{(p)}$. The message from variable node $s_{d,n}^{(p)}$ to factor node $f_{d,n}^{(p)}$ is
\begin{equation}
\begin{aligned}
\nu_{s_{d,n}^{(p)} \rightarrow f_{d,n}^{(p)}}(s_{d,n}^{(p)}) &\propto \nu_{d_{d,n}^{(p)} \rightarrow s_{d,n}^{(p)}}\nu_{d_{d,n+1}^{(p)} \rightarrow s_{d,n}^{(p)}}\\
&=\stackrel{\leftharpoonup}{\pi}_{n}^{(p)}s_{d,n}^{(p)}+(1-\stackrel{\leftharpoonup}{\pi}_{n}^{(p)})(1-s_{d,n}^{(p)}),
\end{aligned}
\end{equation}
with
\begin{equation}
	\stackrel{\leftharpoonup}{\pi}_{n}^{(p)}=\frac{ \lambda_{n}^{f(p)}\lambda_{n}^{b(p)} }{ (1-\lambda_{n}^{f(p)})(1-\lambda_{n}^{b(p)})+\lambda_{n}^{f(p)}\lambda_{n}^{b(p)} }.
\end{equation}
The message from factor node $f_{d,n}^{(p)}$ to variable node $h_{d,n}^{(p)}$ is
\begin{equation}
\begin{aligned}
&\nu_{f_{d,n}^{(p)}\rightarrow h_{d,n}^{(p)}}(h_{d,n}^{(p)}) \\
&\propto \sum_{s_{d,n}^{(p)}}f_{d,n}^{(p)}(h_{d,n}^{(p)},s_{d,n}^{(p)})\cdot\nu_{s_{d,n}^{(p)} \rightarrow f_{d,n}^{(p)}}\\
&=\stackrel{\leftharpoonup}{\pi}_{n}^{(p)}\mathcal{CN}(h_{d,n}^{(p)};0,(\sigma_{d}^{(p)})^{2})+(1-\stackrel{\leftharpoonup}{\pi}_{n}^{(p)})\delta(h_{d,n}^{(p)}).
\end{aligned}
\end{equation}
The posterior mean and variance can be calculated as
\begin{equation}
h_{B,n}^{post(p)} = \text{E}(h_{d,n}^{(p)}|\bm{h}_{B}^{pri(p)}),
\label{equ:postmea2}
\end{equation}
and
\begin{equation}
v_{B}^{post(p)} = \frac{1}{N}\sum_{n=1}^{N}\text{Var}(h_{d,n}^{(p)}|\bm{h}_{B}^{pri(p)}).
\label{equ:postvar2}
\end{equation}

Then, the mean and variance are updated using (\ref{equ:extmea}) and (\ref{equ:extvar}). The structured Turbo-CS algorithm with Module B realized by Eqns. (\ref{equ:td2dd}) to (\ref{equ:postvar2}) is referred to as structured Turbo-CS with delay support (STCS-DS), summarized in Algorithm \ref{alg3}.
Note that both STCS-FS and STCS-DS are approximate algorithms to exploit the sparsity
of the massive MIMO-OFDM channel. Though, it is difficult to tell which algorithm has better
performance in theory, we will show numerically in the next section that STCS-DS makes more efficient
usage of the delay-domain channel sparsity and hence considerably outperforms STCS-FS.


\section{Performance Comparisons}
\label{section:simcomp}
\subsection{Pilot Design}
\begin{figure}[htbp]
	\centering{}
	\includegraphics[width=3in]{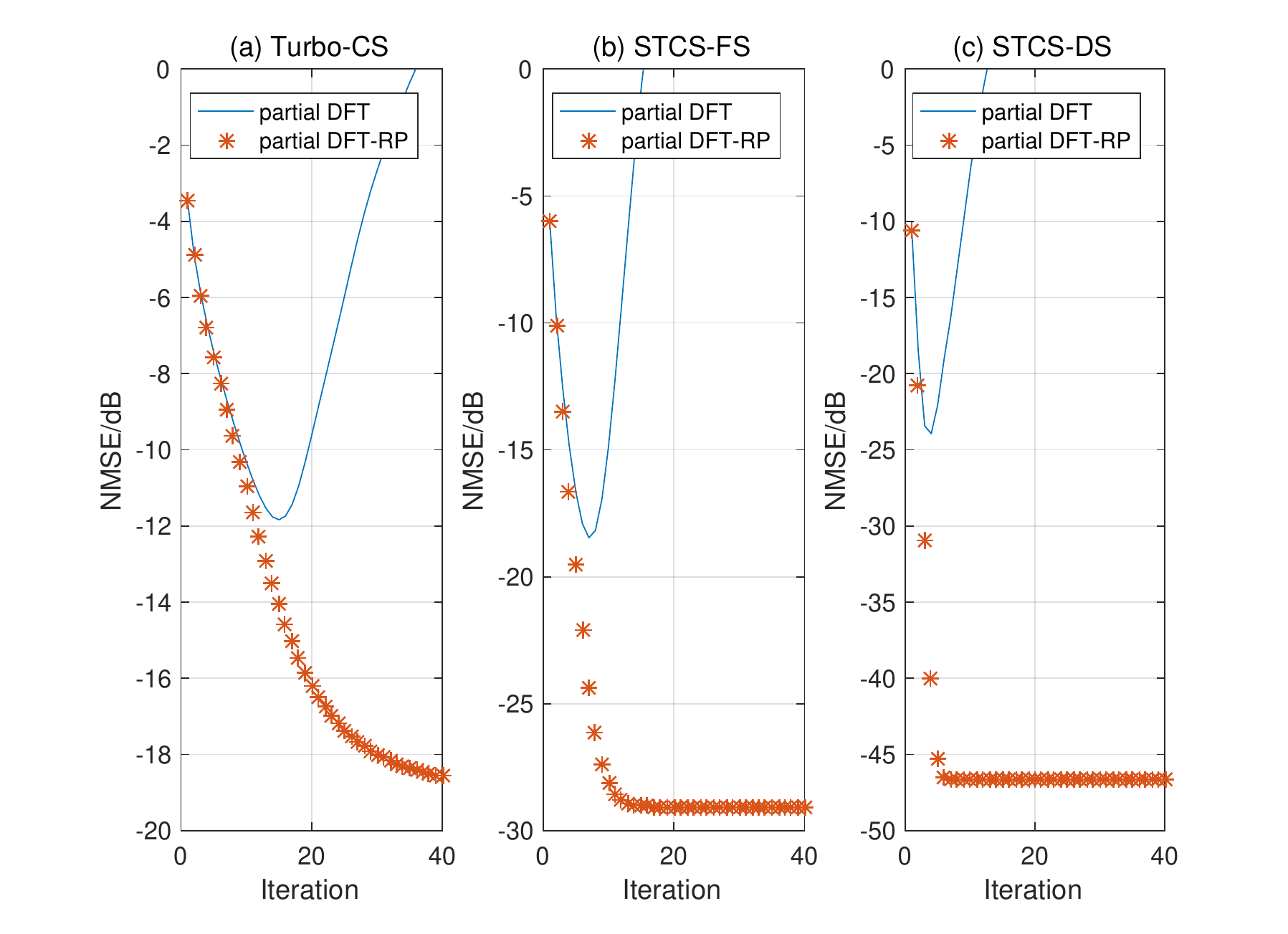}
	\caption{ Algorithm performance comparisons of different sensing matrix under SNR = $30$ dB. The results are averaged by $1000$ realizations.}
	\label{fig:senMatDes}
\end{figure}

The Turbo-CS algorithm and its variants are designed as a low-complexity and near-optimal solution to handle orthogonal measurements, i.e., the sensing matrix is a partial orthogonal matrix. In \cite{Ma2014Turbo}, the sensing matrix is chosen as the partial DFT matrix, which works well for the Turbo-CS algorithm when the unknown variables are i.i.d.. However, as shown in Fig. \ref{fig:senMatDes}, the partial DFT sensing matrix does not work well here, since the support of the channel exhibits a clustered structure, rather than an i.i.d. structure as in \cite{Ma2014Turbo}.

In this work, we decorrelate the sparse signal by using random permutation (RP). The corresponding sensing matrix, referred to as a partial DFT-RP sensing matrix, is given by
\blue{
\begin{equation}
\bm{A} = \bm{S}\bm{F}\bm{R},
\end{equation}
where $\bm{S}$ is a selection matrix consisting of randomly selected and reordered rows of the $N\times N$ identity matrix, and $\bm{R}$ is a random permutation matrix. Then the corresponding pilot matrix is $\bm{X} = \bm{S}\bm{F}\bm{R}\bm{F}$. With such a pilot design, the algorithm only needs to store the permutation orders specified by $\bm{S}$ and $\bm{R}$, rather than to store the whole sensing matrix, which relieves the storage burden at user side. Moreover, the matrix multiplication involving $\bm{A}$ can be realized by the Fast Fourier Transform (FFT) algorithm for complexity reduction.}

In simulation, we consider a massive MIMO-OFDM system with $N = 256$ antennas at BS. Pilot subcarriers are uniformly allocated in the frequency band. The total number of pilot subcarriers is $32$.
\blue{The realizations of the delay taps $\bm{H}_d$ are generated by using the following parameter setting.
 The states $\{s_{d,n}^{(p)}\}$ are generated with transition probability $p_{01}^{(p)} = 1/16$ and $p_{10}^{(p)} = 1/240$.
 Given $\{s_{d,n}^{(p)}\}$, $\bm{H}_d$ is generated by following (11). The maximum delay length $L$ is $16$.
 Once $\bm{H}_d$ is generated, $\bm{H}_f$ can be obtained from \eqref{equ:dftrans}.} The training length is $M = 103 \approx 0.4N$. In Fig. \ref{fig:senMatDes}, Turbo-CS, STCS-FS, and STCS-DS are tested, where the normalized mean square error (NMSE) is defined as $\text{NMSE}=\|\hat{\bm{H}}-\bm{H}\|^{2}_{2}/\|\bm{H}\|^{2}_{2}$.
 From Fig.~\ref{fig:senMatDes}, we see that all the algorithms converge when a partial DFT-RP sensing matrix is used; however, the algorithm diverges when
  a partial DFT sensing matrix is used. The simulation results in later subsections are all based on partial DFT-RP sensing matrices unless otherwise specified.

\subsection{Storage and Computational Complexity}
In original Turbo-CS, the sensing matrix is chosen as a partial DFT matrix, which means the matrix multiplication can be substituted by using FFT. The storage complexity for sensing matrix and computational complexity for each iteration are $\mathcal{O}(1)$ and $\mathcal{O}(N\text{log}N+N)$. For STCS in this paper, we need additional storage $\mathcal{O}(N)$ for permutation matrix and some additional calculations caused by the permutation with computational complexity $\mathcal{O}(N)$. In addition, STCS involves $P$ measurements. Therefore, the proposed STCS-FS and STCS-DS have  per-iteration complexity $\mathcal{O}(PN\text{log}N+PN)$.
 \blue{This per-iteration complexity is lower than that of AMP-NNSPL-DD in \cite{LWJKYH18} $\big($with per-iteration complexity $\mathcal O\left(MPN+NP\log P\right)\big)$. That is, STCS is more
 efficient in both storage and per-iteration complexity than AMP-NNSPL-DD. Later, we
  will further show that STCS also exhibits the fastest convergence rate among all the existing algorithms.} 

\subsection{State Evolution}
The performance of Turbo-CS can be characterized by simple scalar recursions called state evolution \cite{Ma2014Turbo,Ma2015On,xue2017denoising}. We apply a similar technique to STCS by tracking the input variance $\tau_{A}$ and $\tau_{B}$ of Module A and Module B. Specifically, the relation of $\tau_{A}$ and $\tau_{B}$ can be described by $\tau_{B} = f(\tau_{A})$ and $\tau_{A} = g(\tau_{B})$, where $f(\cdot)$ and $g(\cdot)$ correspond to the operations of Module A and B respectively. Then, the fixed point $\tau=g(f(\tau))$ can be used to predict the output mean square error of the STCS algorithm. In this paper, by following \cite{xue2017denoising}, $f(\cdot)$ and $g(\cdot)$ are given by
\begin{equation}
	\tau_{A} = g(\tau_{B}) = \frac{1}{NP}\text{E}\left[ \left\|\bm{D}_{B}(\bm{H}+\tau_{B}\bm{E}) - \bm{H} \right\|_{F}^{2}\right]
	\label{equ:vapri}
\end{equation}
and
\begin{equation}
\tau_{B} = f(\tau_{A}) =  \frac{N}{M}(\tau_{A}+\sigma^{2}) - \tau_{A},
\label{equ:vbpri}
\end{equation}
where $\bm{D}_{B}$ in (\ref{equ:vapri}) is the input output function of Module B with the input $\bm{H}+\tau_{B}\bm{E}$, and each element of matrix $\bm{E}$ obeys a circularly complex Gaussian distribution with zero mean and unit variance. Note that function $\bm{D}_B$ includes not only the structured estimator but also the extrinsic update step. Also note that $g(\cdot)$ in (\ref{equ:vapri}) does not have a simple analytical expression. This function can be numerically evaluated by simulation.

\begin{figure}[htbp]
	\centering{}
	\includegraphics[width=3in]{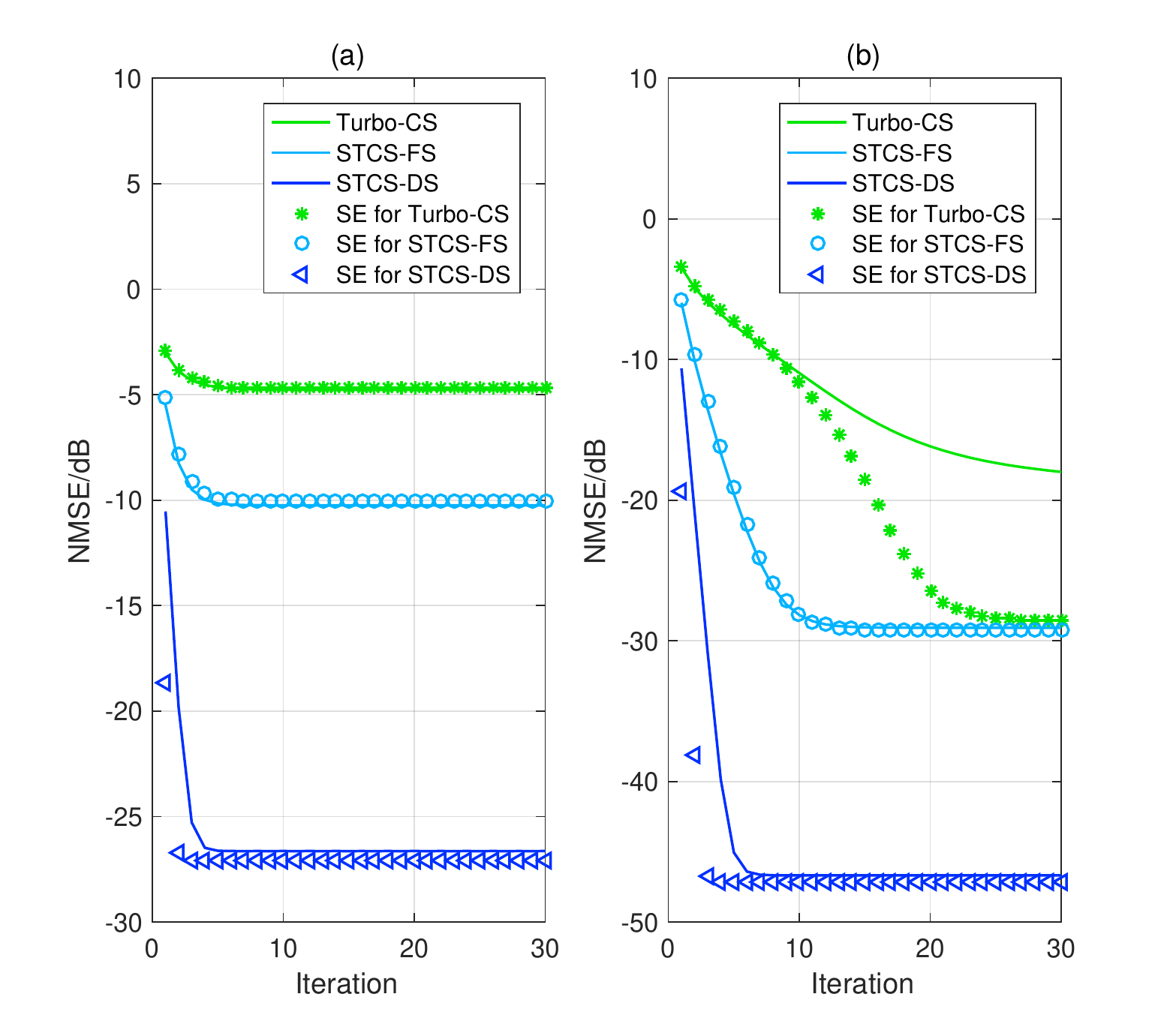}
	\caption{\blue{ Comparison of state evolution and simulation results under SNR = $10$ dB in (a) and $30$ dB in (b). $N = 256$. The results are averaged by $1000$ realizations.}}
	\label{fig:MMVSE}
\end{figure}

Fig. \ref{fig:MMVSE} illustrates the NMSE performances of Turbo-CS and the various STCS-based algorithms proposed in this paper, together with the predictions by the state evolution. In simulation, $N = 256$ and $\text{SNR}=10$ dB and $30$ dB. We see that all the STCS-based algorithms agree well with the state evolution. However, there is a gap for Turbo-CS between simulation and state evolution at $\text{SNR} = 30$ dB. The reason is that the original Turbo-CS algorithm is designed for i.i.d. unknowns, and does not work well for unknowns with clustered sparsity.

\subsection{EM Learning}
The STCS based algorithms require the prior knowledge of the channel distribution. However, the parameters of the channel distribution are usually unknown in practice. In \cite{vila2013expectation}, the expectation maximization (EM) algorithm is combined with the approximate message passing (AMP) algorithm \cite{donoho2009message} to learn the distribution parameters. A similar EM algorithm can be applied to STCS.
For STCS-FS, recall that $\bm{q}_{f} \triangleq [ \lambda_{f}, \sigma_{f}^{(1)}, \cdots, \sigma_{f}^{(P)}, p_{01} ]$ are the parameters of the channel distribution under consideration. Then, in each iteration, the parameters are updated by
\begin{equation}
\bm{q}_{f}^{(t+1)} = \text{arg}\max_{\bm{q}_{f}} \hat{E} \{\text{ln}p(\bm{H}_{f}, \bm{Y}_{f}; \bm{q}_{f})|\bm{Y}_{f}; \bm{q}_{f}^{(t)} \},
\label{equ:EMall}
\end{equation}
where the expection $\hat{E}$ is taken over the output distribution of $\bm{H}_{f}$ in the $t$-th EM iteration. More details of the EM algorithm can be found, e.g., in \cite{vila2013expectation}. Similarly, the EM parameter learning scheme can also be applied to STCS-DS. The parameters of the $p$th delay tap are defined as $\bm{q}_{d}^{(p)} \triangleq [ \lambda_{d}^{(p)}, \sigma_{d}^{(p)}, p_{01}^{(p)}, \gamma_{d}^{(p)} ]$, $\forall p$. Then, in each iteration, the parameters are updated by
\begin{equation}
\bm{q}_{d}^{(p),(t+1)}\!=\!\text{arg}\max_{\bm{q}_{d}^{(p)}} \hat{E} \{\text{ln}p(\bm{h}_{d}^{pri(p)}, \bm{y}_{d}^{(p)}; \bm{q}_{d}^{(p)})|\bm{y}_{d}^{(p)}; \bm{q}_{d}^{(p),(t)} \},
\label{equ:EMall2}
\end{equation}
for all $p$.

\subsection{Noisy Channel Estimation}
\begin{figure*}[htbp]
	\centering{}
	\includegraphics[width=7in]{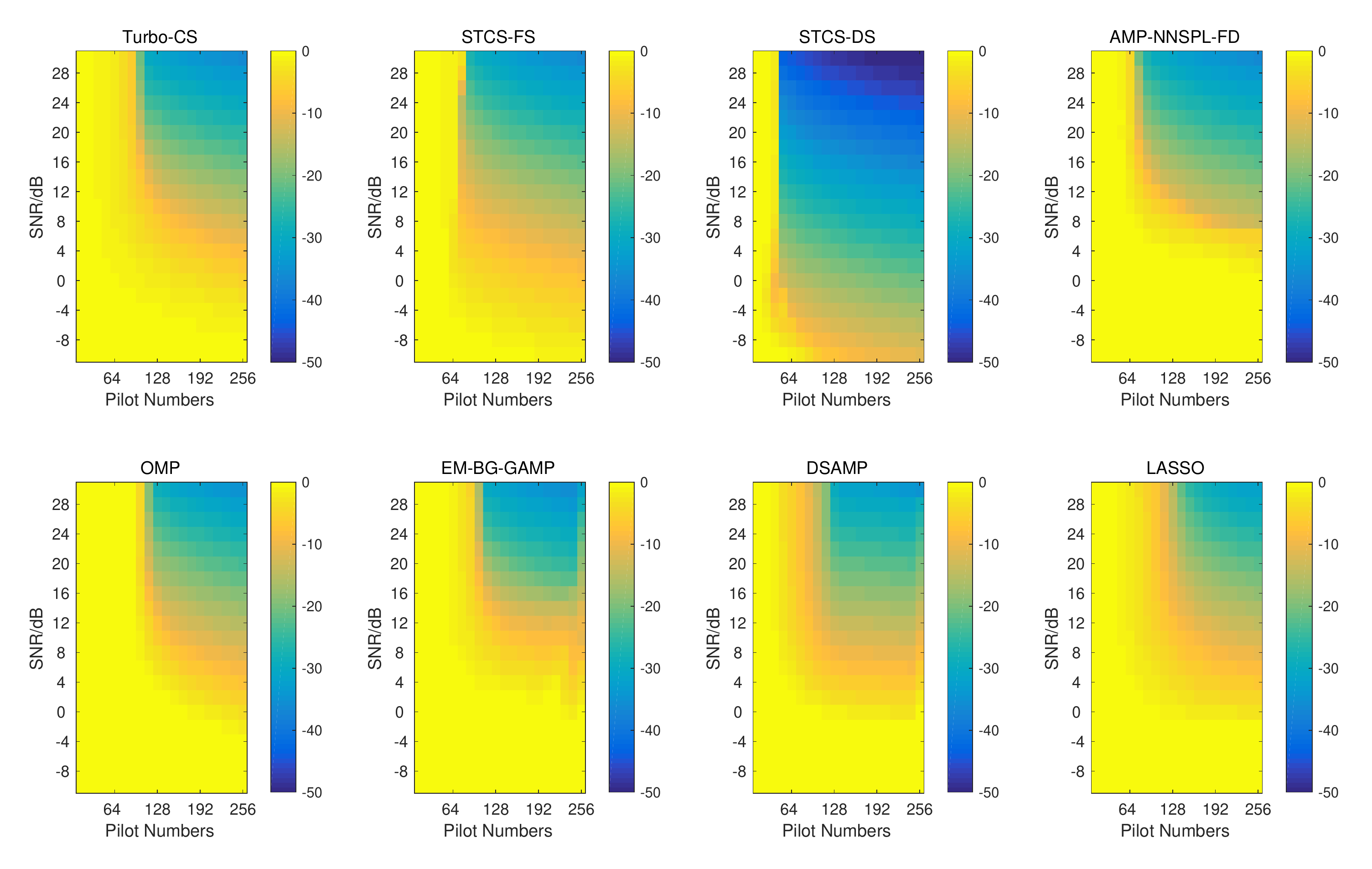}
	\caption{\blue{NMSE performance of various algorithms under $\text{SNR}\in[-10,30]$ dB and pilot numbers $M\in[0.05N,N]$. A $21 \times 20$ grid is constructed for each algorithm with the performance averaged by $100$ independent trials at each grid point.} }
	\label{fig:wideComp}
\end{figure*}

In this subsection, we compare the performance of the proposed STCS-FS and STCS-DS with various baseline algorithms using the channel generated in Subsection A. The parameters of the channel are learned by the EM framework. For frequency support algorithm, the parameters are initialized by $\lambda_{f} = 0.3$, $(\sigma_{f}^{(p)})^{2} = 2N\|\bm{y}_{f}^{(p)}\|^{2}/M\|\bm{A}^{(p)}\|^{2}_{F}$, $\forall p$, and $p_{01} = 0.1$. For delay support  algorithm, the parameters are initialized by $\lambda_{d}^{(p)} = 0.3$, $(\sigma_{d}^{(p)})^{2} = 2N\|\bm{y}_{d}^{(p)}\|^{2}/M\|\bm{A}^{(p)}\|^{2}_{F}$, $p_{01}^{(p)} = 0.1$ and $\gamma_{d}^{(p)} = 0.1$, $\forall p$. In Fig. \ref{fig:wideComp}, we compare the average NMSE performance of OMP \cite{berger2010application}, DSAMP \cite{gao2015spatially}, L1 LASSO \cite{berger2010application}, EM-BG-AMP \cite{vila2013expectation}, Turbo-CS \cite{Ma2014Turbo}, \blue{AMP-NNSPL-FD \cite{LWJKYH18}}, STCS-FS, and STCS-DS under a wide range of SNR and pilot numbers. A $21 \times 20$ grid of each algorithm is constructed from $\text{SNR}\in[-10,30]$ dB and pilot numbers $M\in[0.05N,N]$. The performance is averaged by $100$ independent trials at each grid point. The sensing matrix is always chosen as partial DFT-RP matrix for a fair comparison. From Fig. \ref{fig:wideComp}, we see that the proposed STCS based algorithms, especially STCS-DS, can achieve a considerable gain over all baseline algorithms under various system settings.

\subsection{Test for More Realistic Channel Data}
\blue{We compare the performance of the proposed STCS based algorithms with various baseline algorithms under one realistic channel model: the spatial channel model (SCM) \cite{WinnerScmImplementationIEEETranbst} developed in 3GPP/3GPP2 for low frequency band (less than 6 GHz). The SCM has been widely used to evaluate the channel estimation performance of Massive MIMO-OFDM systems; see, e.g. \cite{XRVL15,liu2016exploiting,chen2017,DLLau18}. In the following, we will use simulations to verify that the proposed STCS algorithms can achieve superior performance over the state-of-the-art baseline algorithms in the realistic channel model under different scenarios, which implies that the proposed probabilistic channel models are flexible and work well for realistic channels.}

\begin{table}[htbp]
	\centering
	\caption{Parameter Settings for the Channel Model}
	\label{tab3}
	\begin{tabular}{cccc}
		\toprule
		\multicolumn{4}{c}{Parameter Settings for the SCM} \\
		\hline
		Parameter name & Value & Parameter name & Value \\
		\hline
		NumBsElements   & 256    & Subcarriers & 512\\
		NumMsElements   & 1      & Subcarrier spacing & 15kHz\\
		CenterFrequency & 2GHz & NumPaths        & 6  \\
		\hline
	\end{tabular}
\end{table}

\begin{figure}[htbp]
	\centering{}
	\includegraphics[width=4in]{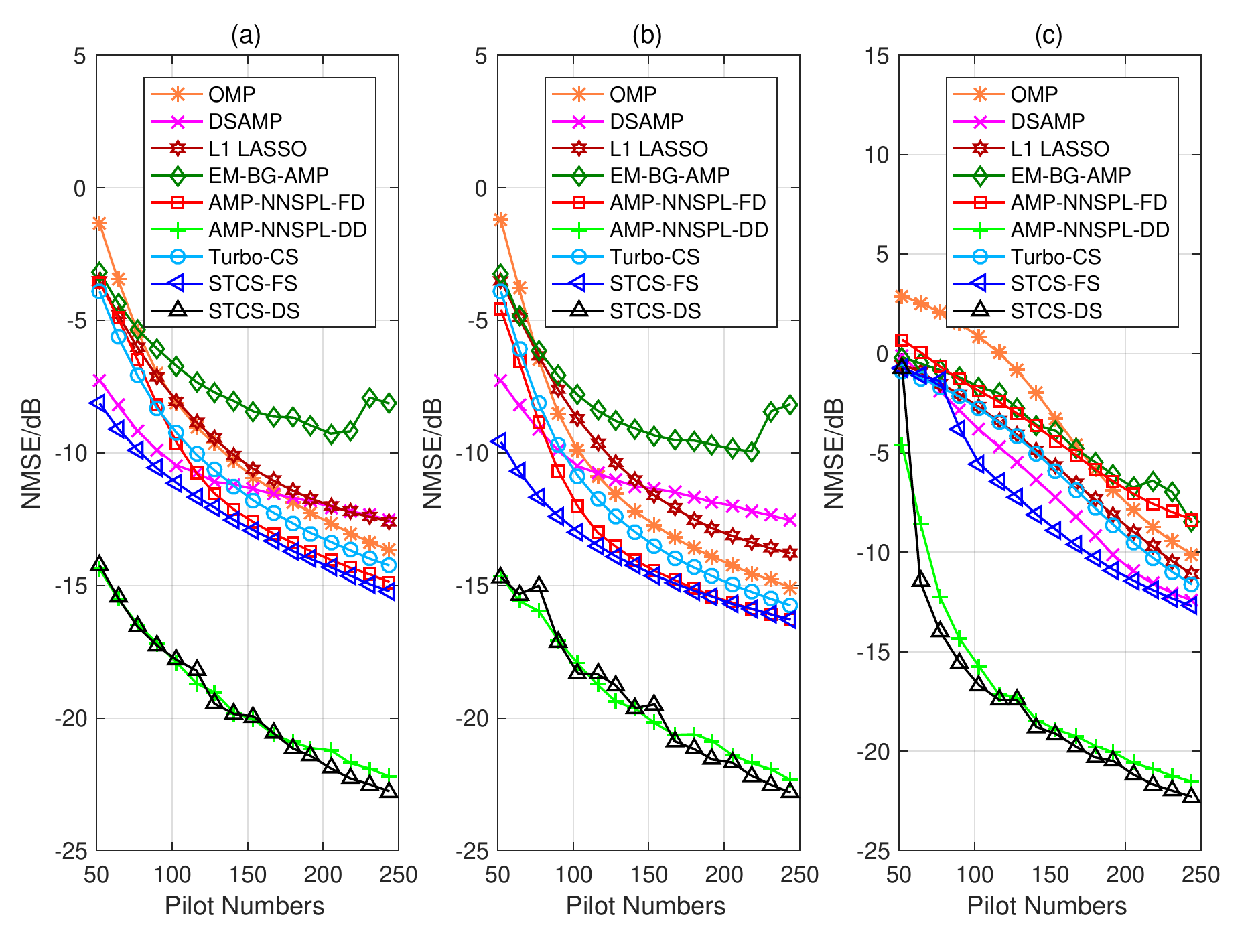}
	\caption{\blue{NMSE of various algorithms versus the number of pilot sequences $M$ under SCM, with $N = 256$ and $\text{SNR} = 10$ dB. The algorithms are tested under different scenario. (a) Urban macro; (b) Suburban macro; (c) Urban micro. The results are averaged by $50$ independent realizations.}}
	\label{fig:Comp_10dB}
\end{figure}

\begin{figure}[htbp]
	\centering{}
	\includegraphics[width=4in]{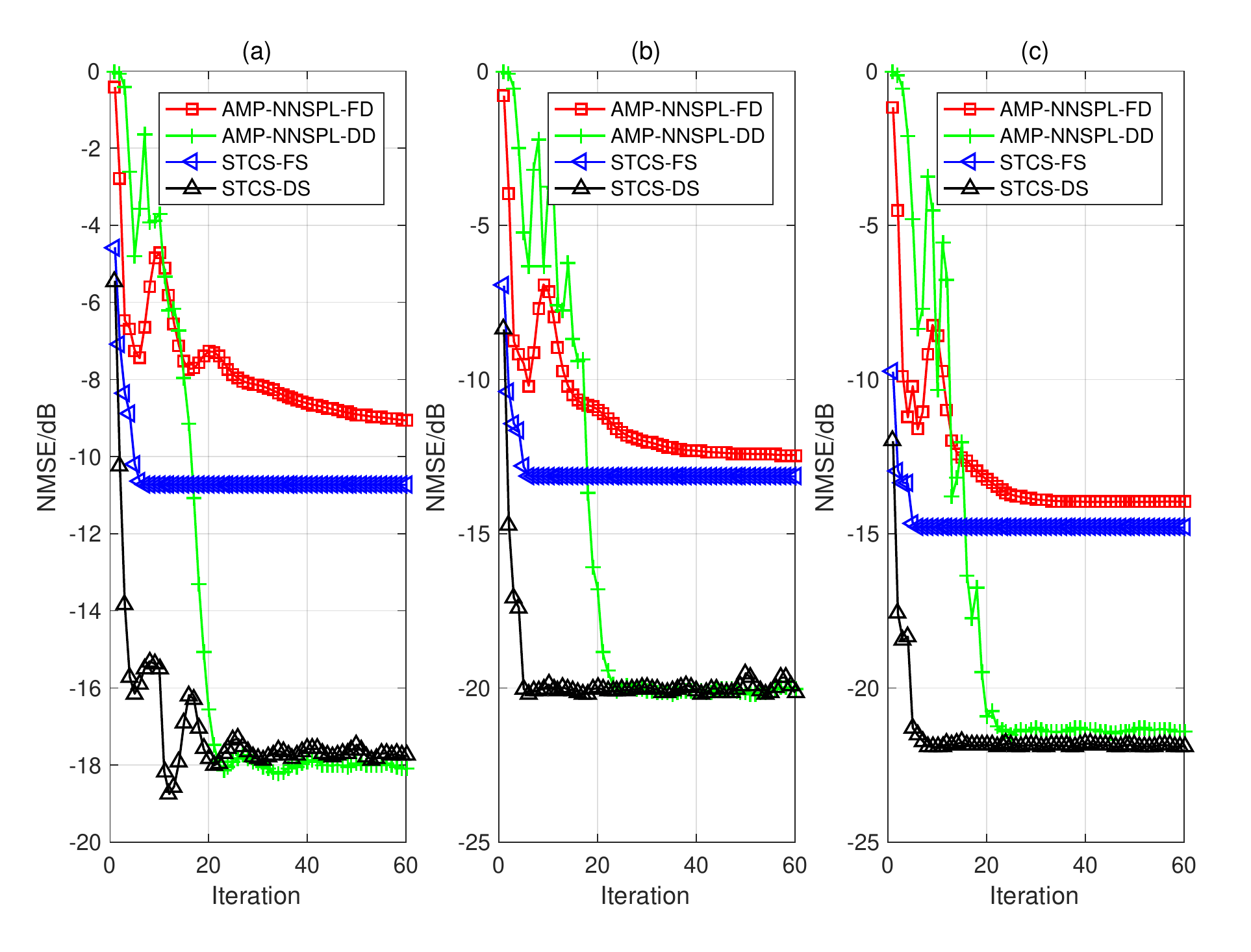}
	\caption{\blue{NMSE versus the iteration number of various algorithms under urban macro scenario at SNR $= 10$ dB, with $N=256$ and $\frac{M}{N}=$ 0.4
in (a), $\frac{M}{N}=$ 0.6 in (b), and $\frac{M}{N}=$ 0.8 in (c). The results are averaged by $50$ independent realizations.}}
	\label{fig:comp.ite}
\end{figure}

The parameters of SCM  used in the simulations are listed in Table \ref{tab3}. \blue{In Fig. \ref{fig:Comp_10dB}, the simulation results are given with $\text{SNR}=10$ dB. 
 We see that the proposed STCS-FS and STCS-DS significantly outperform OMP \cite{berger2010application}, DSAMP \cite{gao2015spatially}, L1 LASSO \cite{berger2010application}, EM-BG-AMP \cite{vila2013expectation}, Turbo-CS \cite{Ma2014Turbo}, and AMP-NNSPL-FD \cite{LWJKYH18} algorithms, while STCS-DS performs slightly better than AMP-NNSPL-DD \cite{LWJKYH18}. This shows the advantage and robustness of the proposed algorithms in practical massive MIMO-OFDM systems. Fig.~\ref{fig:comp.ite} shows the NMSE performances of STCS-based algorithms and the NNSPL-based algorithms \cite{LWJKYH18} as a function of iteration number at SNR$=$ 10 dB, $\frac{M}{N}=$ 0.4 in (a), $\frac{M}{N}=$ 0.6 in (b), and $\frac{M}{N}=$ 0.8 in (c). From Fig.~\ref{fig:comp.ite}, we observe that STCS-based algorithms converge much faster than NNSPL-based algorithms.}
\section{Conclusions}
\label{section:conclu}
In this paper, we apply the structured Turbo-CS framework to improve the estimation accuracy of the massive MIMO-OFDM channel by exploiting its sparsity structure in the angle-frequency domain and angle-delay domains. We show that the proposed STCS based algorithms can be well predicted by the state evolution even for a relatively small $N$. Finally, STCS-FS and STCS-DS are tested for realistic spatial channel models. \blue{We show that the proposed algorithms have much faster convergence speed and achieve competitive NMSE performance under a wide range of simulation settings.} This demonstrates the merit of our channel estimation approach in practical massive MIMO-OFDM systems.


\begin{thebibliography}{10}
\providecommand{\url}[1]{#1}
\csname url@samestyle\endcsname
\providecommand{\newblock}{\relax}
\providecommand{\bibinfo}[2]{#2}
\providecommand{\BIBentrySTDinterwordspacing}{\spaceskip=0pt\relax}
\providecommand{\BIBentryALTinterwordstretchfactor}{4}
\providecommand{\BIBentryALTinterwordspacing}{\spaceskip=\fontdimen2\font plus
\BIBentryALTinterwordstretchfactor\fontdimen3\font minus
  \fontdimen4\font\relax}
\providecommand{\BIBforeignlanguage}[2]{{%
\expandafter\ifx\csname l@#1\endcsname\relax
\typeout{** WARNING: IEEEtran.bst: No hyphenation pattern has been}%
\typeout{** loaded for the language `#1'. Using the pattern for}%
\typeout{** the default language instead.}%
\else
\language=\csname l@#1\endcsname
\fi
#2}}
\providecommand{\BIBdecl}{\relax}
\BIBdecl

\bibitem{chen2017massive}
L.~Chen and X.~Yuan, ``Massive {MIMO}-{OFDM} channel estimation via structured
  turbo compressed sensing,'' in \emph{Proc. IEEE Int. Conf. on Commun. (ICC)},
  2018.

\bibitem{lu2014overview}
L.~Lu, G.~Y. Li, A.~L. Swindlehurst, A.~Ashikhmin, and R.~Zhang, ``An overview
  of massive {MIMO}: Benefits and challenges,'' \emph{IEEE J. Sel. Topics in
  Signal Processing}, vol.~8, no.~5, pp. 742--758, Oct. 2014.

\bibitem{bolcskei2006mimo}
H.~Bolcskei, ``{MIMO}-{OFDM} wireless systems: basics, perspectives, and
  challenges,'' \emph{IEEE Wireless Commun.}, vol.~13, no.~4, pp. 31--37, Aug.
  2006.

\bibitem{wang2014cellular}
C.-X. Wang, F.~Haider, X.~Gao, X.-H. You, Y.~Yang, D.~Yuan, H.~Aggoune,
  H.~Haas, S.~Fletcher, and E.~Hepsaydir, ``Cellular architecture and key
  technologies for 5{G} wireless communication networks,'' \emph{IEEE Commun.
  Magazine}, vol.~52, no.~2, pp. 122--130, Feb. 2014.

\bibitem{yuan2015fundamental}
X.~Yuan, C.~Fan, and Y.~Zhang, ``Fundamental limits of training-based multiuser
  {MIMO} systems,'' \emph{arXiv preprint arXiv:1511.08977}, 2015.

\bibitem{hassibi2003much}
B.~Hassibi and B.~Hochwald, ``How much training is needed in multiple-antenna
  wireless links?'' \emph{IEEE Trans. Info. Theory}, vol.~49, no.~4, pp.
  951--963, 2003.

\bibitem{Gao_TCOM2016_csMIMO}
Z.~Gao, L.~Dai, W.~Dai, B.~Shim, and Z.~Wang, ``Structured compressive
  sensing-based spatio-temporal joint channel estimation for {FDD} massive
  {MIMO},'' \emph{IEEE Trans. Commun.}, vol.~64, no.~2, pp. 601--617, Feb.
  2016.

\bibitem{bajwa2010compressed}
W.~U. Bajwa, J.~Haupt, A.~M. Sayeed, and R.~Nowak, ``Compressed channel
  sensing: A new approach to estimating sparse multipath channels,''
  \emph{Proceedings of the IEEE}, vol.~98, no.~6, pp. 1058--1076, Jun. 2010.

\bibitem{berger2010application}
C.~R. Berger, Z.~Wang, J.~Huang, and S.~Zhou, ``Application of compressive
  sensing to sparse channel estimation,'' \emph{IEEE Commun. Magazine},
  vol.~48, no.~11, Nov. 2010.

\bibitem{tropp2007signal}
J.~A. Tropp and A.~C. Gilbert, ``Signal recovery from random measurements via
  orthogonal matching pursuit,'' \emph{IEEE Trans. Inf. theory}, vol.~53,
  no.~12, pp. 4655--4666, Dec. 2007.

\bibitem{liu2016exploiting}
A.~Liu, V.~K. Lau, and W.~Dai, ``Exploiting burst-sparsity in massive {MIMO}
  with partial channel support information,'' \emph{IEEE Trans. Wireless
  Commun.}, vol.~15, no.~11, pp. 7820--7830, Nov. 2016.

\bibitem{XRVL15}
X.~Rao and V.~Lau, ``Compressive sensing with prior support quality information
  and application to massive {MIMO} channel estimation with temporal
  correlation,'' \emph{IEEE Trans. Signal Processing}, vol.~63, no.~18, pp.
  4914--4924, Sep. 2015.

\bibitem{HLDavid17}
Y.~Han, J.~Lee, and D.~J. Love, ``Compressed sensing-aided downlink channel
  training for {FDD} massive {MIMO} systems,'' \emph{IEEE Trans. Commun.},
  vol.~65, no.~7, pp. 2852--2862, Jul. 2017.

\bibitem{gao2015spatially}
Z.~Gao, L.~Dai, Z.~Wang, and S.~Chen, ``Spatially common sparsity based
  adaptive channel estimation and feedback for {FDD} massive {MIMO},''
  \emph{IEEE Trans. Signal Processing}, vol.~63, no.~23, pp. 6169--6183, Dec.
  2015.

\bibitem{donoho2009message}
D.~L. Donoho, A.~Maleki, and A.~Montanari, ``Message-passing algorithms for
  compressed sensing,'' \emph{Proceedings of the National Academy of Sciences},
  vol. 106, no.~45, pp. 18\,914--18\,919, Nov. 2009.

\bibitem{vila2013expectation}
J.~P. Vila and P.~Schniter, ``Expectation-maximization {G}aussian-mixture
  approximate message passing,'' \emph{IEEE Trans. Signal Processing}, vol.~61,
  no.~19, pp. 4658--4672, Oct. 2013.

\bibitem{Ma2014Turbo}
J.~Ma, X.~Yuan, and L.~Ping, ``Turbo compressed sensing with partial {DFT}
  sensing matrix,'' \emph{IEEE Signal Processing Letters}, vol.~22, no.~2, pp.
  158--161, Feb. 2015.

\bibitem{Ma2015On}
------, ``On the performance of turbo signal recovery with partial {DFT}
  sensing matrices,'' \emph{IEEE Signal Processing Letters}, vol.~22, no.~10,
  pp. 1580--1584, Oct. 2015.

\bibitem{xue2017denoising}
Z.~Xue, J.~Ma, and X.~Yuan, ``Denoising-based turbo compressed sensing,''
  \emph{IEEE Access}, vol.~5, pp. 7193--7204, Apr. 2017.

\bibitem{chen2017}
L.~Chen, A.~Liu, and X.~Yuan, ``Structured turbo compressed sensing for massive
  {MIMO} channel estimation using a {M}arkov prior,'' \emph{IEEE Trans. Veh.
  Technology}, vol.~67, no.~5, May 2018.

\bibitem{MSRJr18}
J.~Mo, P.~Schniter, and R.~W.~H. Jr., ``Channel estimation in broadband
  millimeter wave {MIMO} systems with few-bit {ADCs},'' \emph{IEEE Trans.
  Signal Processing}, vol.~66, no.~5, pp. 1141--1154, Mar. 2018.

\bibitem{ziniel2013dynamic}
J.~Ziniel and P.~Schniter, ``Dynamic compressive sensing of time-varying
  signals via approximate message passing,'' \emph{IEEE Trans. Signal
  Processing}, vol.~61, no.~21, pp. 5270--5284, Nov. 2013.

\bibitem{LWJKYH18}
X.~Lin, S.~Wu, C.~Jiang, L.~Kuang, J.~Yan, and L.~Hanzo, ``Estimation of
  broadband multiuser millimeter-wave massive {MIMO-OFDM} channels by
  exploiting their sparse structure,'' \emph{IEEE Trans. Wireless Commun.},
  vol.~17, no.~6, pp. 3959--3973, Jun. 2018.

\bibitem{MWKHL16}
X.~Meng, S.~Wu, L.~Kuang, D.~Huang, and J.~Lu, ``Approximate message passing
  with nearest neighbor sparsity pattern learning,'' \emph{arXiv:1601.00543v1}.

\bibitem{LWKNMJ17}
X.~Lin, S.~Wu, L.~Kuang, Z.~Ni, X.~Meng, and C.~Jiang, ``Estimation of sparse
  massive {MIMO-OFDM} channels with approximately common support,'' \emph{IEEE
  Commun. Lett.}, vol.~21, no.~5, pp. 1179--1182, May 2017.

\bibitem{WNMK16}
S.~Wu, Z.~Ni, X.~Meng, and L.~Kuang, ``Block expectation propagation for
  downlink channel estimation in massive {MIMO} systems,'' \emph{IEEE Commun.
  Lett.}, vol.~20, no.~11, pp. 2225--2228, Nov. 2016.

\bibitem{Tse:2005:FWC:1111206}
D.~Tse and P.~Viswanath, \emph{Fundamentals of Wireless Communication}.\hskip
  1em plus 0.5em minus 0.4em\relax New York, NY, USA: Cambridge University
  Press, 2005.

\bibitem{WinnerScmImplementationIEEETranbst}
\BIBentryALTinterwordspacing
J.~Salo, G.~{Del Galdo}, J.~Salmi, P.~Kyösti, M.~Milojevic, D.~Laselva, and
  C.~Schneider. (2005, Jan.) {MATLAB} implementation of the {3GPP Spatial
  Channel Model (3GPP TR 25.996)}. [Online]. Available:
  \url{http://www.tkk.fi/Units/Radio/scm/}
\BIBentrySTDinterwordspacing

\bibitem{barhumi2003optimal}
I.~Barhumi, G.~Leus, and M.~Moonen, ``Optimal training design for {MIMO} {OFDM}
  systems in mobile wireless channels,'' \emph{IEEE Trans. Signal Processing},
  vol.~51, no.~6, pp. 1615--1624, Jun. 2003.

\bibitem{dai2013spectrally}
L.~Dai, Z.~Wang, and Z.~Yang, ``Spectrally efficient time-frequency training
  {OFDM} for mobile large-scale {MIMO} systems,'' \emph{IEEE J. Selected Areas
  in Commun.}, vol.~31, no.~2, pp. 251--263, Feb. 2013.

\bibitem{berrou1996near}
C.~Berrou and A.~Glavieux, ``Near optimum error correcting coding and decoding:
  Turbo-codes,'' \emph{IEEE Trans. Commun.}, vol.~44, no.~10, pp. 1261--1271,
  Oct. 1996.

\bibitem{ma2017orthogonal}
J.~Ma and L.~Ping, ``Orthogonal {AMP},'' \emph{IEEE Access}, vol.~5, pp.
  2020--2033, Jan. 2017.

\bibitem{DLLau18}
J.~Dai, A.~Liu, and V.~Lau, ``{FDD} massive {MIMO} channel estimation with
  arbitary 2{D}-array geometry,'' \emph{IEEE Trans. Signal Processing},
  vol.~66, no.~10, pp. 2584--2599, 2018.

\end{thebibliography}

\end{document}